\documentclass[journal]{IEEEtran}
\ifCLASSINFOpdf
\else
   \usepackage[dvipdfmx]{graphicx}
\fi
\usepackage{amsmath}
\newtheorem{theorem}{Theorem}
\newtheorem{remark}{Remark}
\newtheorem{lemma}{Lemma}
\newtheorem{definition}{Definition}
\newtheorem{assumption}{Assumption}
\newtheorem{corollary}{Corollary}

\begin{document}
%
\title{On Second-Moment Stability of Discrete-Time Linear Systems with
General Stochastic Dynamics}
%
%
%

\author{Yohei~Hosoe 
        and~Tomomichi~Hagiwara 
%
\thanks{This work was
supported by JSPS KAKENHI Grant Number 17K14700.}
\thanks{Y.~Hosoe and T.~Hagiwara are with the Department of Electrical Engineering, 
        Kyoto University, Nishikyo-ku, Kyoto 615-8510, Japan
        (e-mail: hosoe@kuee.kyoto-u.ac.jp).}
}

\maketitle

\begin{abstract}
This paper provides a new unified framework for second-moment stability of
 discrete-time linear systems with stochastic dynamics.
Relations of notions of second-moment stability are studied
 for the systems with general stochastic dynamics, and associated
 Lyapunov inequalities are derived.
Any type of stochastic process can be dealt with as a special case in our framework for determining
 system dynamics, and our results together with
 assumptions (i.e., restrictions) on
 the process immediately lead us to stability conditions for the
 corresponding special stochastic systems.
As a demonstration of usefulness of such a framework, three selected
 applications are also provided.
\end{abstract}

\begin{IEEEkeywords}
Discrete-time linear systems, stochastic dynamics, stability analysis,
 Lyapunov inequalities, LMIs.
\end{IEEEkeywords}

%
\IEEEpeerreviewmaketitle

\section{Introduction}

Randomness is a common concept in many fields, with
which various kinds of phenomenon are interpreted and evaluated.
Examples can be readily found such as packet interarrival times in networks
\cite{Paxson-IEEETN95},
failure occurrences in distributed systems \cite{Finkelstein-book}
and chance of precipitation in weather forecast \cite{Kalnay-book}.
The systems having this kind of randomness (more precisely, the systems
whose underlying randomness is regarded as essential) are 
called stochastic systems.
This paper focuses on randomness of dynamics rather than that of input
for discrete-time linear systems, for which internal stability and associated Lyapunov
inequalities are discussed toward future control applications.

The systems with stochastic dynamics are called random dynamical
systems in the field of analytical dynamics \cite{Arnold-book}.
The arguments in this paper begin with the most general 
stochastic dynamics for discrete-time linear systems, whose system class
is completely
consistent with the discrete-time linear case of random dynamical systems.
One of the motivations of our dealing with such systems is that 
the system class is considered to be compatible with
sequential Monte Carlo methods such as
the ensemble Kalman filter \cite{Evensen-EnKF,Evensen-book} and
the Gaussian mixture filter \cite{kotecha2003gaussian,stordal2011bridging}.
%
Parameters and their distributions in systems can be sequentially estimated by those
methods, and their mixture with theory for
stochastic systems is expected to open a new frontier of control with sequential
learning.
This paper has the role of providing a way to give a guarantee on control
performance (in stability) for the stochastic systems from the
theoretical viewpoint.
In particular, several notions of second-moment stability are
introduced, and their relationships are discussed for the stochastic
systems.
Then, associated Lyapunov inequalities are derived, with which the
convergence rate of the second moment of system state can be evaluated.

While systems with general stochastic dynamics are relatively complicated to deal
with for control even in our discrete-time linear case and few results have been reported, 
some subclasses are well studied, each of which has formed an
independent research field in the control society.
One of the famous subclasses is Markov jump systems \cite{Costa-book}.
Although the dynamics of a standard Markov jump system is described with
a finite-mode Markov chain, the earlier study \cite{Costa-TAC14} succeeded in alleviating
this part of assumption so that a more general stationary
Markov process can be dealt with in stability analysis.
Another noteworthy subclass is the systems with white parameters
\cite{Koning1992whitepara}, which we call the
systems with dynamics determined by an independent and identically
distributed (i.i.d.) process \cite{Hosoe-TAC18,Hosoe-TAC19}.
This subclass further involves systems with state-multiplicative noise
\cite{Boyd-book,Gershon-Auto08}.

Each of the above subclasses has an independent research history in the
control society.
At least in analysis of second-moment stability, however, 
our results turn out to deal with all the above systems in a
unified framework, which facilitates the study of
their relationships and generalizations drastically. 
For example, there is a difference between the above Markov and i.i.d. cases
that the former (i.e., Markov case) assumed the essential boundedness of
coefficient random matrices (depending on a Markov process) for derivation
of a Lyapunov inequality in the earlier study while the latter did not; a random matrix
depending on a standard Markov chain (i.e., the coefficients of a
standard Markov jump system) obviously satisfies this assumption.
Because of this difference, the results in the i.i.d.\ case were not
covered by those in the Markov case, even though i.i.d.\ processes are
a special case of Markov processes; indeed, the Lyapunov inequality in  
\cite{Costa-TAC14} does not readily reduce to that in \cite{Hosoe-TAC19} 
even when the processes are restricted to i.i.d.\ type.
By using the results in this paper, we can easily clarify the essential reason of this
difference.
In addition to such an academic investigation, our results can be used also
for generalization of earlier results as already stated, e.g., so that periodically
stationary (i.e., periodically distributed) processes can be dealt with.
Some associated applications will be provided later as a demonstration
of powerfulness of our new framework.

The purposes of this paper are summarized as follows:
(i) complete systematization of theory for second-moment stability of
discrete-time linear systems with general stochastic dynamics, (ii)
clarification of relationships among some subclasses of stochastic
systems in stability analysis, and (iii) generalization of selected
earlier results.
The purposes (i) and (ii) are related with academic significance, and the
rest is with usefulness of the proposed framework.


This paper is organized as follows.
Section~\ref{sc:sys} describes discrete-time linear systems with
dynamics determined by a general stochastic process,
and states the treatment of the initial condition for the systems 
associated with the underlying processes.
Since the processes determining stochastic dynamics in this paper are
general, we constantly use conditional expectations, with which some
readers might be less familiar.
Hence, the section also makes a brief preliminary for conditional
expectations.
After these preliminaries, five notions of second-moment
stability are introduced, whose relations are discussed in
Section~\ref{sc:relations} as a part of main results in this paper.
The relations in the most general case of
systems are first discussed, and then, 
further relations are discussed under an assumption on the systems to
have a sort of 'time-invariance' property.
Since exponential stability, which is one of the above five stability
notions, is compatible with stability analysis based on Lyapunov
inequalities, Section~\ref{sc:Lyap} derives the Lyapunov inequalities
characterizing it without loss of generality.
In particular, we drives two types of Lyapunov inequalities, one of which
is for the general systems and the other is for the systems having
essentially bounded coefficient matrices.
These results also become a key in clarifying the reason of the difference between 
the conventional results for Markov and i.i.d.\ cases stated above.
As a demonstration of powerfulness of our results,
Section~\ref{sc:app} provides some selected applications, which not only
clarify relationships of earlier and our results but also
generalize the former in a very simple fashion; our results
together with 
additional assumptions on the systems immediately lead us to the
corresponding Lyapunov inequality conditions (including conventional ones).
The framework proposed in this paper can unify all the results about second-moment stability of
discrete-time linear systems having stochastic dynamics, and is expected to 
facilitate the studies in this field drastically.

We use the following notation in this paper.
The set of real numbers,
that of positive real numbers,
that of integers
and that of non-negative integers are denoted by
${\bf R}$, ${\bf R}_+$, ${\bf Z}$ and ${\bf N}_0$, respectively.
Subsets of ${\bf Z}$ are defined as
${\bf Z}_+(t):=[t, \infty) \cap {\bf Z}$ and ${\bf Z}_-(t):=(-\infty,
t]\cap {\bf Z}$ for $t\in {\bf Z}$.
The set of $n$-dimensional real
column vectors and that of $m\times n$ real matrices
are denoted by ${\bf R}^n$ and ${\bf R}^{m\times n}$, respectively.
The set of $n\times n$ symmetric matrices and that of 
$n\times n$ positive definite matrices are denoted by 
${\bf S}^{n\times n}$ and ${\bf S}^{n\times n}_{+}$, respectively.
The Borel $\sigma$-algebra on the set $(\cdot)$ is denoted by ${\cal B}(\cdot)$.
The maximum 
singular value is denoted by $\sigma_{\rm max}(\cdot)$.
The Euclidean norm is denoted by $||(\cdot)||$.
For random variables $s_1$ and $s_2$, the expectation of $s_1$ and the
conditional expectation of $s_1$ given $s_2$ are denoted by $E[s_1]$ and
$E[s_1 |\, s_2]$, respectively;
this notation is used also for random matrices.


\section{Discrete-Time Linear Systems with General Stochastic
 Dynamics and Second-Moment Stability}
\label{sc:sys}

Let $(\Omega,{\cal F},P)$ be a complete probability space, where
$\Omega$, ${\cal F}$ and $P$ are a sample space, a $\sigma$-algebra and 
a probability measure, respectively.
All the random variables and processes in this paper will be defined on
this common probability space.
That is, for a set ${\boldsymbol X}$, an ${\boldsymbol X}$-valued random variable $X_0$ is defined as
a mapping $X_0: (\Omega, {\cal F}) \rightarrow ({\boldsymbol X}, {\cal
B}({\boldsymbol X}))$; we describe this mapping also as
$X_0: \Omega \rightarrow {\boldsymbol X}$ for short.
Similarly, an ${\boldsymbol X}$-valued stochastic process $X=(X_k)_{k\in
{\boldsymbol T}}$ on the
set ${\boldsymbol T}$ of time instants is defined as a mapping $X: \Omega
\rightarrow {\boldsymbol X}^{\boldsymbol T}$ (i.e., $X: (\Omega, {\cal F})
\rightarrow ({\boldsymbol X}^{\boldsymbol T}, {\cal
B}({\boldsymbol X}^{\boldsymbol T}))$), where 
${\boldsymbol X}^{\boldsymbol T}$ 
is the set of all the possible 
${\boldsymbol X}$-valued functions of $k\in {\boldsymbol T}$ that map ${\boldsymbol
T}$ to ${\boldsymbol X}$.
For details of the terms about probability theory, see
\cite{Knill-book,Klenke-book} and other sophisticated books.

With the above probability space, this section first describes
discrete-time linear systems with general stochastic dynamics.
Then, several definitions for second-moment stability are given.

\subsection{Discrete-Time Linear Systems with General Stochastic
 Dynamics}

Let us consider the ${\bf R}^{n\times n}$-valued (i.e., matrix-valued)
stochastic process $\tilde{A}=(\tilde{A}_{t})_{t\in {\bf Z}}: \Omega \rightarrow ({\bf
R}^{n\times n})^{\bf Z}$ defined on $(\Omega,{\cal F},P)$,
and the associated discrete-time linear system
\begin{equation}
x_{k+1} = \tilde{A}_{k} x_{k}
\label{eq:fr-sys-orig}
\end{equation}
with the finite-dimensional state $x_k$,
where $k$ is
the discrete time (which is supposed to go forward).
It is obvious that the
above equation describes the most general 
discrete-time linear finite-dimensional (input-free) systems with stochastic dynamics,
if no restrictions are imposed on $(\Omega,{\cal F},P)$ and $\tilde{A}$.

For convenience in discussing technical results, we next
introduce an alternative representation of such systems without causing
any loss of generality in the system description.
To this end, we first consider the column expansion of
$\tilde{A}_k$ for each $k$, and denote it by $\xi_k\in{\bf R}^{n^2}$.
Then, it is obvious that
$\tilde{A}_k=A(\xi_k)$ by introducing the time-invariant mapping
$A(\cdot)$ in an obvious fashion.
This observation immediately implies that confining the system description to
\begin{equation}
 x_{k+1} = A(\xi_{k}) x_{k},
\label{eq:fr-sys}
\end{equation}
where $\xi_k\in{\bf R}^Z$ for each $k$, does not lead to any loss of
generality compared with (\ref{eq:fr-sys-orig}),
as long as the classes of the probability space $(\Omega,{\cal F},P)$,
the integer $Z\in{\bf N}$, the
stochastic process
$\xi=(\xi_k)_{k\in {\bf Z}}: \Omega \rightarrow ({\bf R}^Z)^{\bf Z}$,
and the Borel-measurable matrix-valued function
$A: {\bf R}^Z \rightarrow {\bf R}^{n\times n}$
are arbitrary.
After discussing in the following subsection
how the initial condition at a given initial time instant
$k\in{\bf Z}$ should be handled for these systems,
this paper discusses for the first time the stability problems of
the general stochastic system in the form (\ref{eq:fr-sys}).

\subsection{Treatment of the Initial Condition}

This paper basically assumes that we are given the initial time instant
$k_0\in{\bf Z}$, and is interested in studying the behavior of the state
$x_k$ of (\ref{eq:fr-sys-orig}) or (\ref{eq:fr-sys})
for $k\in{\bf Z}_+(k_0)$, even though $k_0$ is eventually assumed
to be arbitrary so that stability of these systems can be
defined appropriately and then studied thoroughly.
Hence, each time $k_0$ is fixed,
we assume that the initial state $x_{k_0}\in{\bf R}^n$ is given,
and we regard it as a deterministic vector.
This initial state alone, however, is not enough as the initial condition
of these systems when we aim at discussing their behavior for
$k\in{\bf Z}_+(k_0)$.  This is because these systems are associated with
the stochastic process $\tilde{A}$, and the behavior of $x_k$ for
$k\in{\bf Z}_+(k_0)$ depends on the (conditional) distribution of
$(\tilde{A}_k)_{k\in{\bf Z}_+(k_0)}$ given all the information available
at time $k_0$.  This implies that the initial condition of these systems
consists not only of the initial state $x_{k_0}$ but also of the initial
condition of the stochastic process $\tilde{A}$ at $k_0$.
For example, when $\tilde{A}$ is a Markov process in (\ref{eq:fr-sys-orig}),
its initial condition is nothing but the conditional distribution of
$\tilde{A}_{k_0}$ given $\tilde{A}_{k_0-1}$,
where $\tilde{A}_{k_0-1}$ is viewed as a deterministic matrix.
Similarly, when $\xi$ is a Markov process in (\ref{eq:fr-sys}),
its initial condition is nothing but
the conditional distribution of $\xi_{k_0}$
given $\xi_{k_0-1}$, where $\xi_{k_0-1}$ is viewed as a deterministic vector.

For adequately dealing with the initial condition at $k_0$ of the general
stochastic process $\xi$ in (\ref{eq:fr-sys}),
let us introduce its subsequences 
$\xi^{k+}=(\xi_{t})_{t\in {\bf Z}_+(k)}: \Omega \rightarrow ({\bf R}^Z)^{{\bf Z}_+(k)}$
and 
$\xi^{k-}=(\xi_{t})_{t\in {\bf Z}_-(k)}: \Omega \rightarrow ({\bf R}^Z)^{{\bf
Z}_-(k)}$ for each $k\in {\bf Z}$.
The intuitive interpretation of 
such partitioning of the stochastic process $\xi$ is that
$\xi^{(k-1)-}$ for each $k\in{\bf Z}$ can be regarded as the
(possibly redundant) information that is sufficient for determining 
the distribution of $\xi^{k+}$
(more precisely, its conditional distribution available at time $k$).
In particular, when $k$ equals the initial time instant $k_0$ of
the system (\ref{eq:fr-sys}), the associated $\xi^{(k_0-1)-}$ can be
regarded 
as determining the initial condition of the (future) stochastic process
$\xi^{k_0+}$, which together with the initial state $x_{k_0}$ determines
the behavior of $x_k$ for $k\in{\bf Z}_+(k_0)$.
Hence, this paper assumes that  $\xi^{(k_0-1)-}$ viewed as a (past) deterministic vector series
is given as information determining the initial condition of
the stochastic process $\xi$ at $k_0$,
together with the initial state $x_{k_0}$.

In our later discussions on stability, the initial state vector $x_{k_0}$ will
be treated as being arbitrary in ${\bf R}^n$.
Similarly, the initial condition of $\xi^{k_0+}$ at $k_0$ will be treated
as being arbitrary by regarding $\xi^{(k_0-1)-}$ as being arbitrary
(past) vector series in its support,
which we denote by $\widehat{\boldsymbol \Xi}_0$.

\subsection{Preliminaries about Conditional Expectation}

In this paper, we deal with several notions of second-moment stability \cite{Kozin-Auto69}.
Roughly speaking, second-moment stability is the concept about the
convergence of (or the existence of a certain uniform bound for) the second moment
of $\|x_{k}\|$ with respect to $k\in {\bf Z}_+(k_0)$.
As is clear from our treatment of the initial condition
for (\ref{eq:fr-sys}), the second moment of $\|x_{k}\|$ should precisely
refer to
the conditional expectation of $\|x_{k}\|^2$ given $\xi^{(k_0-1)-}$.
When we wish to be clear that $\xi^{(k_0-1)-}$ is viewed as
a deterministic series in this context, we could instead introduce the
notation $\widehat{\xi}^{(k_0-1)-}$ to denote the path of the
(past) {\it stochastic process} $\xi^{(k_0-1)-}$ up to time $k_0-1$;
in this context, the initial
condition of $\xi^{k_0+}$ at $k_0$ could more precisely be written
as the equation $\xi^{(k_0-1)-}=\widehat{\xi}^{(k_0-1)-}$.
Under this notational standpoint,
the second moment can be written as the conditional expectation
 $E[\|x_{k}\|^2|\, \xi^{(k_0-1)-} = \widehat{\xi}^{(k_0-1)-}]$.
Since the conditional expectations of other quantities are also
handled repeatedly,
we introduce the shorthand notation
\begin{align}
E_0[(\cdot)]:=E[(\cdot)|\, \xi^{(k_0-1)-} = \widehat{\xi}^{(k_0-1)-}].
\label{eq:def-E0}
 \end{align}
Stability will be defined later through this notion of conditional expectations.

For each $k \in {\bf Z}_+(k_0)$ together with the underlying
initial condition $\xi^{(k_0-1)-}=\widehat{\xi}^{(k_0-1)-}$,
we further introduce 
the $\sigma$-algebra ${\cal F}_{k}$
generated by $\xi_{k_0}, \xi_{k_0+1}, \ldots, \xi_{k}$ (under the
underlying initial condition
$\widehat{\xi}^{(k_0-1)-}\in\widehat{\boldsymbol \Xi}_0$), and another associated conditional expectation
\begin{align}
E_0[(\cdot)|\,{\cal F}_{k}]:=E[(\cdot)|\, \xi^{(k_0-1)-} =
 \widehat{\xi}^{(k_0-1)-}, \xi_{k_0}, \xi_{k_0+1}, \ldots, \xi_{k}].
\end{align}
By definition, this conditional expectation can be seen as a
random variable depending only on $\xi_{k_0}, \xi_{k_0+1}, \ldots,
\xi_{k}$ (and the initial condition $\xi^{(k_0-1)-}=\widehat{\xi}^{(k_0-1)-}$).
In addition,
${\cal F}_{k}\subset {\cal F}_{k+1}$ and
${\cal F}_{k} \subset {\cal F}$ for each $k \in {\bf Z}_+(k_0)$.
Hence, $({\cal F}_{k})_{{k}\in {\bf Z}_+(k_0)}$ is a filtration
for each $k_0\in {\bf Z}$ and every $\widehat{\xi}^{(k_0-1)-}\in \widehat{\boldsymbol \Xi}_0$.
In this paper, we repeatedly use conditional expectations.
To facilitate understanding of the associated arguments, a brief review
of some basic properties of the conditional expectation is given
before proceeding to stability definitions.

For $k_1\geq k_0$,
let $f:({\bf R}^Z)^{{\bf Z}_+(k_1)} \rightarrow {\bf R}$ be a
Borel-measurable function.
If $f(\xi^{k_1+})\geq 0$ holds almost surely (a.s.) for the underlying
initial condition at $k_0$, then $E_0[f(\xi^{k_1+})| {\cal
F}_{k_2}]\geq 0\
{\rm a.s.}$ regardless of $k_2 \in {\bf Z}_+(k_0)$.
If $f(\xi^{k_1+})$ is Lebesgue integrable (i.e., its expectation is finite),
\begin{align}
& E_0[E_0[f(\xi^{k_1+})| {\cal F}_{k_3}]| {\cal F}_{k_2}]=E_0[f(\xi^{k_1+})| {\cal
 F}_{k_2}]\ \ (k_3\geq k_2).\label{eq:con-exp-2}
\end{align}
Let $g:({\bf R}^Z)^{{\bf Z}_-(k_1-1)} \rightarrow {\bf R}$ 
be also a Borel-measurable function.
If $g(\xi^{(k_1-1)-})f(\xi^{k_1+})$ is Lebesgue integrable,
\begin{align}
& E_0[g(\xi^{(k_1-1)-})f(\xi^{k_1+})| {\cal F}_{k_2}]
\notag \\
=&
g(\xi^{(k_1-1)-})E_0[f(\xi^{k_1+})| {\cal F}_{k_2}]\ \ (k_2\geq k_1-1).\label{eq:con-exp-3}
\end{align}
In particular, when $f(\xi^{k_1+})=1$,
\begin{align}
 & E_0[g(\xi^{(k_1-1)-})| {\cal F}_{k_2}]=g(\xi^{(k_1-1)-})\ \ (k_2\geq k_1-1).\label{eq:con-exp-1}
\end{align}
These properties are a natural consequence from those of the standard conditional expectation.

\subsection{Stability Notions}
\label{ssc:stab}


To define second-moment stability, we introduce the following assumption on system (\ref{eq:fr-sys}).

\begin{assumption}
\label{as:bound}
For each $k_0\in {\bf Z}$,
there exists $M_1=M_1(k_0)\in {\bf R}_+$ such that
%
\begin{align}
&
E_0\left[A_{ij}(\xi_{k_0})^2\right] \leq M_1(k_0)\notag \\
&
 (\forall i, j = 1,\ldots,n; \forall \widehat{\xi}^{(k_0-1)-}\in \widehat{\boldsymbol \Xi}_0),
\label{eq:as-bound}
\end{align}
where $A_{ij}(\xi_{k_0})$ denotes the $(i,j)$-entry of $A(\xi_{k_0})$.
\end{assumption}

Since 
\begin{align}
&
\sigma_{\max}(A(\xi_{k_0}))^2\leq
\sum_{i=1}^n \sum_{j=1}^n A_{ij}(\xi_{k_0})^2 \label{eq:msv-ent}
\end{align}
regardless of $\xi_{k_0}$,
this assumption implies the existence of a $k_0$-dependent upper bound of
$E_0[\sigma_{\max}(A(\xi_{k_0}))^2]$.
The assumption is a minimal requirement for 
dealing with the second moment $E_0\left[\|x_{k}\|^2\right]$
for each initial time instant $k_0$ and every $k\in{\bf Z}_+(k_0)$,
as can be confirmed in the following lemma.

\begin{lemma}
\label{lm:bound-iff}
For system (\ref{eq:fr-sys}),
the following two conditions are equivalent.
\begin{enumerate}
\item
Assumption~\ref{as:bound} is satisfied.
\item  For each $k_0\in {\bf Z}$ and every $k\in {\bf Z}_+(k_0)$,
there exists $M_2=M_2(k,k_0)\in {\bf R}_+$ such that
\begin{align}
& E_0\left[\|x_{k}\|^2\right] \leq M_2(k,k_0)\|x_{k_0}\|^2 \notag \\
&
(\forall x_{k_0} \in {\bf R}^n; \forall \widehat{\xi}^{(k_0-1)-}\in \widehat{\boldsymbol \Xi}_0). \label{eq:well-defined-2nd-moment}
\end{align}
That is, the second moment $E_0\left[\|x_{k}\|^2\right]$ is bounded for
	   each $k_0$ and every $k$.
%
%
\end{enumerate}
\end{lemma}

\begin{IEEEproof}
1$\Rightarrow$2:
Fix $k_0$ and $\widehat{\xi}^{(k_0-1)-}$, and take an arbitrary $k\in{\bf Z}_+(k_0)$.
We proceed with the proof by tentatively relating this $k$ with $k_0$ in
(\ref{eq:as-bound})
so that we could evaluate the conditional expectation
of $A_{ij}(\xi_k)^2$ or $\sigma_{\max}(A(\xi_k))^2$ in an appropriate sense
(while maintaining the overall standpoint that $k_0$ always refers to the
one we fixed at the beginning of the proof {\it except for
the specific tentative treatment here}).
More precisely, we give a restatement
(in terms of $k\in{\bf Z}_+(k_0)$ rather than $k_0$ that we fixed)
of what (\ref{eq:as-bound}) implies with respect to $A_{ij}(\xi_k)$
when Assumption~\ref{as:bound} is satisfied.
First, the assumption obviously implies that the conditional expectation
of $E_0[A_{ij}(\xi_k)^2|{\cal F}_{k-1}]$ given $\widehat{\xi}^{(k_0-1)-}$
is well-defined for $i,j=1,\dots,n$ regardless of
$\xi_{k_0},\dots,\xi_{k-1}$ such that the vector series
$\widehat{\xi}^{(k_0-1)-}$, $\xi_{k_0},\dots,\xi_{k-1}$ belongs
to the support of $\xi^{(k-1)-}$
(where $E_0[\cdot]$ is with respect to $k_0$ and $\widehat{\xi}^{(k_0-1)-}$
that we fixed at the beginning of the proof).
Hence, by (\ref{eq:msv-ent}),
$\alpha(k)=n^2 M_1(k)$ satisfies
\begin{align}
&
E_0\left[\sigma_{\max}(A(\xi_{k}))^2| {\cal F}_{k-1}\right]\leq \alpha(k)\
 {\rm a.s.}
\label{eq:as-bound-equiv2}
\end{align}
%
for each $k\in{\bf Z}_+(k_0)$.
This, together with (\ref{eq:con-exp-2}) and (\ref{eq:con-exp-3}), leads us to
\begin{align}
&
E_0\left[\|x_{k+1}\|^2\right] \notag\\
=&
E_0\left[ \|A(\xi_{k}) x_{k}\|^2 \right] \notag\\
\leq&
E_0\left[ \sigma_{\max}(A(\xi_{k}))^2 \|x_{k}\|^2\right] \notag\\
=&
E_0\left[ E_0\left[\sigma_{\max}(A(\xi_{k}))^2 \|x_{k}\|^2| {\cal F}_{k-1}\right] \right] \notag\\
=&
E_0\left[ E_0\left[\sigma_{\max}(A(\xi_{k}))^2| {\cal F}_{k-1}\right] \|x_{k}\|^2 \right] \notag\\
\leq&
\alpha(k)E_0\left[ \|x_{k}\|^2 \right]\notag
\\
&
(\forall k \in {\bf Z}_+(k_0); \forall x_{k_0} \in {\bf R}^n).
\end{align}
%
%
A recursive use of this inequality leads to
(\ref{eq:well-defined-2nd-moment}) with
\begin{align}
& M_2(k,k_0):=\alpha(k_0)\alpha(k_0+1)\cdots \alpha(k-1).
\end{align}
%
%
This completes the proof.

2$\Rightarrow$1:
By taking $k=k_0+1$, the inequality in
(\ref{eq:well-defined-2nd-moment}) leads to
\begin{align}
x_{k_0}^T E_0[A(\xi_{k_0})^T A(\xi_{k_0})] x_{k_0}
&=
 E_0\left[\|A(\xi_{k_0})x_{k_0}\|^2\right] \notag \\
&\leq M_2(k_0+1,k_0) \|x_{k_0}\|^2.
\end{align}%
Since this inequality holds for all $x_{k_0} \in {\bf R}^n$,
there exists $M_1=M_1(k_0)$ satisfying $0\leq M_1(k_0) \leq M_2(k_0+1,k_0)$ and
 (\ref{eq:as-bound}).
This completes the proof.
\end{IEEEproof}


For system (\ref{eq:fr-sys}) satisfying
Assumption~\ref{as:bound}, we first define the most basic notion of second-moment
stability as follows.

\begin{definition}[Stability]
\label{df:stab}
The system (\ref{eq:fr-sys}) (satisfying
Assumption~\ref{as:bound}) is said to be stable in the second moment
 if for each $\epsilon\in {\bf R}_+$ and every $k_0\in {\bf Z}$, there exists
 $\delta=\delta(\epsilon,k_0)$ such that
\begin{align}
&
\|x_{k_0}\|^2 \leq \delta(\epsilon,k_0) \Rightarrow
E_0[\|x_{k}\|^2] \leq \epsilon\notag \\
& 
(\forall k \in {\bf Z}_+(k_0); \forall \widehat{\xi}^{(k_0-1)-}\in \widehat{\boldsymbol \Xi}_0).\label{eq:stab-def}
\end{align}
\end{definition}

Note that the inequality (\ref{eq:well-defined-2nd-moment})
resulting from Assumption~\ref{as:bound} ensures that it is indeed meaningful
to refer to $E_0[\|x_{k}\|^2]$ in the above definition
(but (\ref{eq:well-defined-2nd-moment}) itself does not
immediately imply the second uniform inequality in $k$
in the above definition).

Uniform stability is further defined as
follows.

\begin{definition}[Uniform Stability]
\label{df:unif-stab}
The system (\ref{eq:fr-sys}) is said to be uniformly stable in the second moment
 if for each $\epsilon\in {\bf R}_+$, there exists
 $\delta=\delta(\epsilon)$ such that
\begin{align}
&
\|x_{k_0}\|^2 \leq \delta(\epsilon) \Rightarrow
E_0[\|x_{k}\|^2] \leq \epsilon\notag \\
&
(\forall k\in
{\bf Z}_+(k_0); \forall \widehat{\xi}^{(k_0-1)-}\in \widehat{\boldsymbol
 \Xi}_0; \forall k_0 \in {\bf Z}).\label{eq:unif-stab-def}
\end{align}
\end{definition}

The above two stability notions are about the existence of a certain
(i.e, $k$-independent) upper bound uniform in $k$ for the second moment
$E_0[\|x_{k}\|^2]$.
The following three notions are about the convergence of the second
moment to 0.

\begin{definition}[Asymptotic Stability]
\label{df:asym-stab}
The system (\ref{eq:fr-sys}) is said to be asymptotically stable in the second moment
 if it is stable in the second moment and for each $k_0\in {\bf Z}$,
\begin{align}
&
E_0[\|x_{k}\|^2] \rightarrow 0\ {\rm as}\ k\rightarrow \infty\ \
 (\forall x_{k_0} \in {\bf R}^n; \forall \widehat{\xi}^{(k_0-1)-}\in \widehat{\boldsymbol \Xi}_0).\label{eq:asy-def}
\end{align}
\end{definition}

\begin{definition}[Uniform Asymptotic Stability]
\label{df:unif-asym-stab}
The 
system (\ref{eq:fr-sys}) is said to be
uniformly asymptotically stable
 in the second moment if 
 it is uniformly stable in the second moment and
\begin{align}
&
E_0[\|x_{k}\|^2] \rightarrow 0\ 
{\rm as}\ 
k\rightarrow \infty\ \ (\forall x_{k_0} \in {\bf R}^n;
\forall \widehat{\xi}^{(k_0-1)-}\in
 \widehat{\boldsymbol \Xi}_0),\notag \\
&
{\rm uniformly~in}\ k_0\in {\bf Z}.\label{eq:unif-asy-def}
\end{align}
What the latter condition (\ref{eq:unif-asy-def}) precisely means is that 
for each $\epsilon\in {\bf R}_+$,
there exists $k_0$-independent $K=K(\epsilon)\in {\bf N}_0$ such that
\begin{align}
&
E_0[\|x_{k}\|^2] \leq \epsilon\|x_{k_0}\|^2\notag \\
&
(\forall k \in {\bf Z}_+(k_0+K(\epsilon)); \forall x_{k_0} \in {\bf
 R}^n; \forall \widehat{\xi}^{(k_0-1)-}\in
 \widehat{\boldsymbol \Xi}_0; \forall k_0 \in {\bf Z}).
\label{eq:unif-asym-equiv}
\end{align}
\end{definition}

\begin{definition}[Exponential Stability]
\label{df:expo-stab}
The system (\ref{eq:fr-sys}) is said to be exponentially stable in the second moment
 if there exist $a \in {\bf R}_+$ and $\lambda \in
 (0,1)$ such that
\begin{align}
&
 E_0[||x_{k}||^2] \leq a ||x_{k_0}||^2 \lambda^{2(k-k_0)}\notag
\\
&
(\forall k \in {\bf Z}_+(k_0); \forall x_{k_0} \in {\bf R}^n; \forall \widehat{\xi}^{(k_0-1)-}\in \widehat{\boldsymbol \Xi}_0; \forall
k_0\in {\bf Z}).
\label{eq:exp-def}
\end{align}
\end{definition}

One of the main purposes of this paper is to discuss the relationships
among these stability notions for system (\ref{eq:fr-sys}).


\begin{remark}
Second-moment stability is also called mean square stability \cite{Kozin-Auto69} in some
literature.
Hence, for instance, the stability notion defined in
Definition~\ref{df:asym-stab} corresponds to asymptotic mean square
stability, which might be more familiar to some readers.
Such paraphrases could facilitate understanding of the relationship
between our study and other earlier results.
\end{remark}

\section{Relations of Stability Notions}
\label{sc:relations}

\subsection{Relations for general stochastic systems}

\begin{figure*}[t]
\centering
  \includegraphics[width=0.9\linewidth]{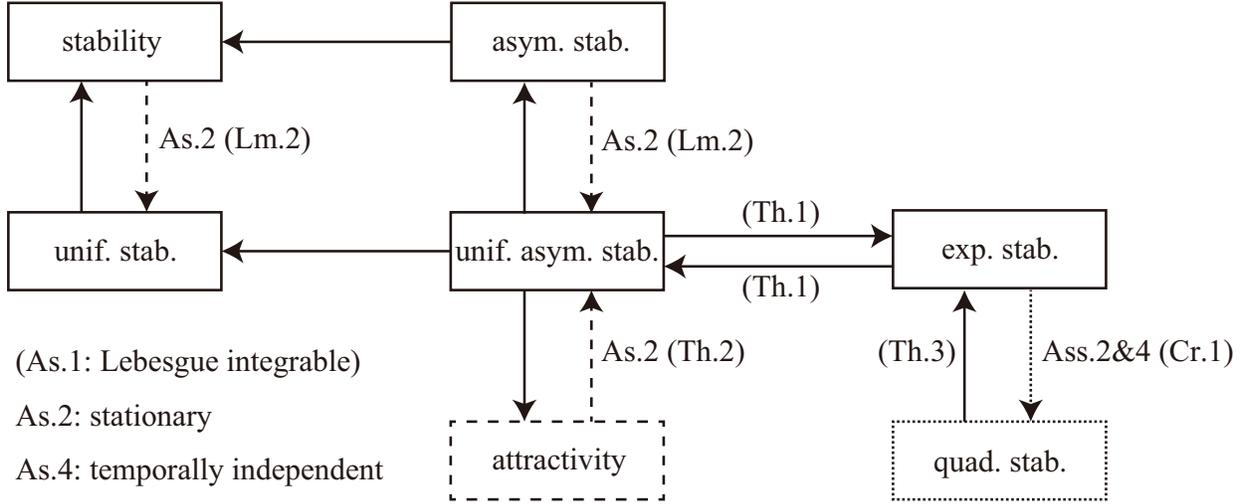}
  \caption{Relations of notions of second-moment stability for system
 (\ref{eq:fr-sys}) under Assumption~\ref{as:bound}.}
\label{fig:gen_relations}
\end{figure*}

This section first discusses the relations of the five stability notions
introduced above for the most general case of 
discrete-time linear systems with stochastic dynamics (i.e., only under Assumption~\ref{as:bound}).
To state the conclusion in advance, the relations shown with solid arrows in
Fig.~\ref{fig:gen_relations} hold for the system (\ref{eq:fr-sys})
under Assumption~\ref{as:bound}, where
all the arrows are in the direction from a strong stability notion to a weak
stability notion.
If the two notions are linked with two arrows with different directions,
the notions are equivalent under the corresponding assumptions (some
assumptions are additionally introduced later).
By definitions of stability notions, the relations described with
solid arrows are trivial except the equivalence between uniform asymptotic stability
and exponential stability (the relations about attractivity and quadratic stability will
be discussed separately after introducing additional assumptions).
Hence, we here focus only on the non-trivial equivalence, and show 
the following theorem for the general stochastic systems.

\begin{theorem}
\label{th:eqv-asym-expo}
Suppose the system (\ref{eq:fr-sys}) satisfies
 Assumption~\ref{as:bound}.
The following two conditions are equivalent.
\begin{enumerate}
 \item 
The system is uniformly asymptotically stable in the second moment.
\item
	 The system is exponentially stable in the second moment.
\end{enumerate}
\end{theorem}

\begin{IEEEproof}

2$\Rightarrow$1:
This assertion is almost obvious; it follows from (\ref{eq:exp-def}) and
 $0<\lambda<1$ that the system is uniformly stable, and for each $\epsilon\in {\bf R}_+$,
there exists $K=K(\epsilon)\in {\bf N}_0$ satisfying (\ref{eq:unif-asym-equiv}).


\medskip
1$\Rightarrow$2:
Linearity of the system (\ref{eq:fr-sys}) frequently used in
this part of the proof is not explicitly referred to so as not to make
 the arguments verbose.
The initial step for the proof of this assertion is similar to
Theorem~1 in \cite{Hosoe-TAC19}.
That is, we first introduce the decomposition
\begin{align}
&
x_{k_0}=\beta_{k_0} \sum_{i=1}^n a_{k_0 i} \sigma_{k_0 i} e^{(i)}
\end{align}
with the non-negative scalars $\beta_{k_0}$, 
$a_{k_0 i}\ (i=1,\ldots,n)$ satisfying $\sum_{i=1}^n a_{k_0 i}=1$,
the integers $\sigma_{k_0 i}\in \{-1,1\}\ (i=1,\ldots,n)$
and the standard basis vectors $e^{(i)}\ (i=1,\ldots,n)$ for the
$n$-dimensional Euclidean space.
By definition, we have
\begin{align}
\|x_{k_0}\|^2
&=
\beta_{k_0}^2(a_{k_0 1}^2+\ldots +a_{k_0 n}^2)
\geq
\beta_{k_0}^2/n.\label{eq:th1-pf-x0low}
\end{align}
Associated with this decomposition of $x_{k_0}$, we can also decompose 
the corresponding $x_{k}$ as
\begin{align}
&
x_{k}=\beta_{k_0} \sum_{i=1}^n a_{k_0 i} \sigma_{k_0 i} x^{(i)}_{k},
\label{eq:xkk0-deco}
\end{align}
where $x^{(i)}_{k}$ is the state at $k$ for the initial state
$x_{k_0}=e^{(i)}$.
It follows from  (\ref{eq:unif-asym-equiv}) that 
there exists $K\in {\bf N}_0$ such that
\begin{align}
&
E_0[\|x^{(i)}_{k}\|^2]\leq 1/(2n^2)\notag\\
&(i=1,\ldots,n; \forall k\in {\bf Z}_+(k_0+K);
\forall \widehat{\xi}^{(k_0-1)-}\in
 \widehat{\boldsymbol \Xi}_0; \forall k_0 \in {\bf Z}).
\end{align}
Then, we have
\begin{align}
&E_0[\|x_{k}\|^2]\notag\\
=&
\beta_{k_0}^2
E_0\left[
\left\|\sum_{i=1}^n a_{k_0 i} \sigma_{k_0 i} x^{(i)}_{k}\right\|^2\right] \notag \\
\leq&
\beta_{k_0}^2 
E_0\left[
\sum_{i=1}^n a_{k_0 i} \|\sigma_{k_0 i} x^{(i)}_{k}\|^2\right] \notag \\
=&
\beta_{k_0}^2 \sum_{i=1}^n a_{k_0 i} E_0[\|x^{(i)}_{k}\|^2] \notag \\
\leq&
\beta_{k_0}^2/(2n^2)\notag\\
&(\forall k\in {\bf Z}_+(k_0+K); \forall x_{k_0}\in {\bf R}^n;
\forall \widehat{\xi}^{(k_0-1)-}\in
 \widehat{\boldsymbol \Xi}_0;
 \forall k_0 \in {\bf Z}),\label{eq:th1-pf-xkup}
\end{align}
where the first inequality follows from Jensen's inequality.
Hence, it follows from (\ref{eq:th1-pf-x0low}) and (\ref{eq:th1-pf-xkup})
that
\begin{align}
&
E_0[\|x_{K+k_0}\|^2]\leq\|x_{k_0}\|^2/2\notag \\
&
(\forall x_{k_0}\in {\bf R}^n;
\forall \widehat{\xi}^{(k_0-1)-}\in
 \widehat{\boldsymbol \Xi}_0;
 \forall k_0 \in {\bf Z}) \label{eq:proof-th1-rec-orig}
\end{align}
for the same $K$.
Since this inequality holds for each $k_0 \in {\bf Z}$ and every $\widehat{\xi}^{(k_0-1)-}$
 belonging to the support $\widehat{\boldsymbol \Xi}_0$ of
  ${\xi}^{(k_0-1)-}$ (recall the arguments about the treatment of $k_0$ at the beginning of the proof of
 Lemma~\ref{lm:bound-iff}),
it follows
for each $k\in {\bf Z}_+(k_0)$ and every sample of the series $\xi_{k_0},\ldots,\xi_{k-1}$
 determining $x_k$ that
\begin{align}
&
E[\|x_{k+K}\|^2|\ {\xi}^{(k_0-1)-}=\widehat{\xi}^{(k_0-1)-}, \xi_{k_0},\ldots,\xi_{k-1}]
\leq \|x_{k}\|^2/2 \notag \\
& (\forall x_{k_0}\in {\bf R}^n;
\forall \widehat{\xi}^{(k_0-1)-}\in
 \widehat{\boldsymbol \Xi}_0;
\forall k_0 \in {\bf Z}),
\end{align}
which implies
\begin{align}
&
E_0[\|x_{k+K}\|^2| {\cal F}_{k-1}]
\leq \|x_{k}\|^2/2\ {\rm a.s.}\notag \\
& (\forall
 k\in {\bf Z}_+(k_0); \forall x_{k_0}\in {\bf R}^n;
\forall \widehat{\xi}^{(k_0-1)-}\in
 \widehat{\boldsymbol \Xi}_0;
\forall k_0 \in {\bf Z}).
\end{align}
This together with (\ref{eq:con-exp-2}) further implies
\begin{align}
&
E_0[\|x_{k+K}\|^2]
\leq E_0[\|x_{k}\|^2]/2
\notag \\
&(\forall
 k\in {\bf Z}_+(k_0); \forall x_{k_0}\in {\bf R}^n; 
\forall \widehat{\xi}^{(k_0-1)-}\in
 \widehat{\boldsymbol \Xi}_0;
\forall k_0 \in {\bf Z}).
\label{eq:proof-th1-rec}
\end{align}
For each $k\in  {\bf Z}_+(k_0)$, take $c, j\in
 {\bf N}_0$ such that 
$k=c+jK+k_0\ (0\leq c<K)$.
Then, a recursive use of (\ref{eq:proof-th1-rec}) leads to
\begin{align}
&
E_0[\|x_{k}\|^2]\notag \\
=&E_0[\|x_{c+jK+k_0}\|^2] \notag\\
\leq&
E_0[\|x_{c+k_0}\|^2]/2^j
\notag \\
=&
2^{c/K}E_0[\|x_{c+k_0}\|^2](2^{-1/K})^{(k-k_0)}
\notag \\
\leq&
2E_0[\|x_{c+k_0}\|^2](2^{-1/K})^{(k-k_0)}\notag\\
&(\forall k
 \in {\bf Z}_+(k_0); \forall x_{k_0}\in {\bf R}^n;
\forall \widehat{\xi}^{(k_0-1)-}\in
 \widehat{\boldsymbol \Xi}_0;
 \forall k_0 \in {\bf Z}).
\label{eq:proof-th1-expo}
\end{align}
Here, $x_{c+k_0}$ can be decomposed as (\ref{eq:xkk0-deco}), and thus, satisfies
\begin{align}
&E_0[\|x_{c+k_0}\|^2]
\leq
\beta_{k_0}^2 \sum_{i=1}^n a_{k_0 i} E_0[\|x^{(i)}_{c+k_0}\|^2]\notag\\
&(\forall c\in [0,K); \forall x_{k_0}\in {\bf R}^n;
\forall \widehat{\xi}^{(k_0-1)-}\in
 \widehat{\boldsymbol \Xi}_0;
 \forall k_0 \in {\bf Z}).\label{eq:xck0-deco}
\end{align}
Since the system is uniformly stable (and
 $\|x^{(i)}_{k_0}\|^2=\|e^{(i)}\|^2=1$),
there exists $\epsilon^{\prime}$ such that
\begin{align}
&
E_0[\|x^{(i)}_{c+k_0}\|^2]\leq \epsilon^{\prime}
\notag\\
&(i=1,\ldots,n; \forall c\in [0,K);
\forall \widehat{\xi}^{(k_0-1)-}\in
 \widehat{\boldsymbol \Xi}_0;
 \forall k_0 \in {\bf Z}),
\label{eq:k0-independent}
\end{align}
which together with (\ref{eq:th1-pf-x0low}) and (\ref{eq:xck0-deco}) implies
\begin{align}
&E_0[\|x_{c+k_0}\|^2]
\leq
n \epsilon^{\prime}\|x_{k_0}\|^2\notag\\
&(\forall c\in [0,K); \forall x_{k_0}\in {\bf R}^n;
\forall \widehat{\xi}^{(k_0-1)-}\in
 \widehat{\boldsymbol \Xi}_0;
 \forall k_0 \in {\bf Z}).\label{eq:xck0-bound}
\end{align}
Hence, from (\ref{eq:proof-th1-expo}) and (\ref{eq:xck0-bound}),
we obtain
\begin{align}
&
E_0[\|x_{k}\|^2]
\leq
2n \epsilon^{\prime}\|x_{k_0}\|^2
(2^{-1/K})^{(k-k_0)}\notag \\
&
(\forall k
 \in {\bf Z}_+(k_0); \forall x_{k_0}\in {\bf R}^n;
\forall \widehat{\xi}^{(k_0-1)-}\in
 \widehat{\boldsymbol \Xi}_0;
 \forall k_0 \in {\bf Z}).
\end{align}
This implies
the 
existence of $a=2n \epsilon^{\prime}$ and $\lambda=2^{-1/K}$ such
 that $a\in {\bf R}_+$, $\lambda\in (0,1)$ and (\ref{eq:exp-def}) hold.
This completes the proof.
\end{IEEEproof}

The essential differences between this proof and that for Theorem~1 in
\cite{Hosoe-TAC19}
are the treatment of the conditional expectation from
(\ref{eq:proof-th1-rec-orig}) through (\ref{eq:proof-th1-rec}) and the use of
uniform stability in taking $k_0$-independent $\epsilon^\prime$ in
(\ref{eq:k0-independent}) so that $a$ in (\ref{eq:exp-def}) is $k_0$-independent.
The latter will be related with the discussions on the time-invariance
property of stochastic systems in the following subsection.

\subsection{Time-invariance property of stochastic systems}

The relations described by the solid arrows in Fig.~\ref{fig:gen_relations} are known to
hold also for linear time-varying deterministic systems \cite{Vidyasagar-book}.
In the case of deterministic systems, it is also known that stability
implies uniform stability if the system is time-invariant.
A similar result can be obtained for our stochastic systems by using the following
assumption.
\begin{assumption}
\label{as:stationary}
The stochastic process $\xi$ 
is stationary (in the strict sense); i.e., 
none of the characteristics of $\xi_k$ changes with time $k$.
\end{assumption}

This assumption leads to the following lemma (see the dashed arrows
in Fig.~\ref{fig:gen_relations}), which is
obvious from the definition of each stability notion.
\begin{lemma}
Suppose the system (\ref{eq:fr-sys}) satisfies Assumptions~\ref{as:bound}
 and \ref{as:stationary}.
The system is uniformly stable in the second moment if and only if the
 system is stable in the second moment.
Similarly, the system is uniformly asymptotically stable in the second moment if and only if the
 system is asymptotically stable in the second moment.
\end{lemma}

In addition, Assumption~\ref{as:stationary} and
Theorem~\ref{th:eqv-asym-expo} lead us to the following theorem.
\begin{theorem}
\label{th:attract}
Suppose the system (\ref{eq:fr-sys})
satisfies Assumptions~\ref{as:bound}
 and \ref{as:stationary}.
The 
system is
 uniformly asymptotically stable
 in the second moment if and only if (\ref{eq:unif-asy-def}) holds.
\end{theorem}
\begin{IEEEproof}
Since the necessity assertion is obvious, we here only prove the
 sufficiency assertion, i.e., uniform asymptotic stability being implied only with
 (\ref{eq:unif-asy-def}) (under Assumptions~\ref{as:bound}
 and \ref{as:stationary}).
Since uniform asymptotic stability is implied by
exponential stability under Assumption~\ref{as:bound}
by Theorem~\ref{th:eqv-asym-expo},
it suffices to show that (\ref{eq:unif-asy-def}) implies exponential
stability under Assumption~\ref{as:stationary}.
The key point for showing this claim is to note where we had to use uniform stability (i.e.,
 (\ref{eq:unif-stab-def})) in showing
 exponential stability in the part 1$\Rightarrow$2 of the proof of
 Theorem~\ref{th:eqv-asym-expo}; if we can show it with additional Assumption~\ref{as:stationary} instead
of the uniform stability assumption, then
 the proof is completed.
Regarding the above key point, we readily see that
uniform stability was used in showing the
 existence of the $k_0$-independent constant $\epsilon^\prime$ in
 (\ref{eq:k0-independent}).
Such a $k_0$-independent constant always exists if the system
 satisfies not only Assumption~\ref{as:bound} but
 Assumption~\ref{as:stationary}; indeed,
\begin{align}
\sup_{i\in\{1,\ldots,n\}, c\in \{0,\ldots,K-1\}, \widehat{\xi}^{(k_0-1)-}\in
 \widehat{\boldsymbol \Xi}_0}
E_0[\|x^{(i)}_{c+k_0}\|^2]
\end{align}
is bounded by Assumption~\ref{as:bound} (recall
 Lemma~\ref{lm:bound-iff}) and $k_0$-independent by
 Assumption~\ref{as:stationary}, and taking it
as $\epsilon^\prime$ in (\ref{eq:k0-independent})
is sufficient.
This completes the proof.
\end{IEEEproof}

The property described by (\ref{eq:unif-asy-def}) is called (uniform) attractivity \cite{Vidyasagar-book}.
It is known
in the case of deterministic systems
that asymptotic stability can be ensured only with attractivity if the
system is linear and time-invariant.
Hence, Theorem~\ref{th:attract} corresponds to a stochastic counterpart of
such a result for deterministic systems.
Although a similar result was already shown in
our earlier study \cite{Hosoe-TAC19} for stochastic systems as Corollary~1,
it was assumed throughout the study that $\xi_k$ 
is independent and identically distributed (i.i.d.) with respect to
$k\in {\bf N}_0$ (i.e., $\xi$ is not only stationary but also temporally-independent).
The present Theorem~\ref{th:attract} was derived only with
Assumption~\ref{as:stationary} (in addition to
Assumption~\ref{as:bound}),
and hence, is more general than the corollary in the earlier study.

According to the above arguments, Assumption~\ref{as:stationary} seems to
let the stochastic system (\ref{eq:fr-sys}) have a sort of ``time-invariance'' property.
However, it will be clearer in the next section that exponential stability is not equivalent
to quadratic stability (dealt with, e.g., in \cite{extended-Oliveira-SCL99,Daafouz-SCL01} for
deterministic systems and in \cite{Hosoe-TAC19} for
a special case of stochastic systems) only with
Assumption~\ref{as:stationary} because exponential stability cannot be
characterized by the Lyapunov inequality with a constant (i.e.,
time-invariant deterministic) Lyapunov matrix\footnote{The definition 
of quadratic stability may be arguable for stochastic systems.
For example, the earlier study \cite{boukas1998stochastic} for Markov
jump systems has the stance that quadratic stability is defined with a
mode-dependent (i.e., non-constant) Lyapunov matrix.
However, another earlier study \cite{geromel2006stability} for switched
systems has the stance that quadratic stability is defined with a
mode-independent (i.e., constant) Lyapunov matrix.
Since the $A$ matrix in our system is time-varying, the present paper
has the same stance as the latter study.} in that case;
it is shown in \cite{Hosoe-TAC19} that if $\xi_k$ is i.i.d.\ with respect to $k$
then exponential stability becomes equivalent to quadratic stability, as
is the case with linear time-invariant deterministic systems.
Hence, the complete ``time-invariance'' property does not follow only with
Assumption~\ref{as:stationary}, and other additional assumptions (e.g.,
temporal independence of $\xi$) are
needed depending on the required level of the property.
This is a difference between the case of deterministic systems and that
of stochastic systems.

\section{Lyapunov Inequalities}
\label{sc:Lyap}

Among the five stability notions introduced in Section~\ref{sc:sys},
exponential stability is the most compatible with the 
approach of stability analysis using Lyapunov inequalities.
Hence, we first derive in this section Lyapunov inequalities giving a necessary and sufficient
condition for exponential stability of system (\ref{eq:fr-sys}) only
under Assumption~\ref{as:bound} (i.e., the most general case);
it is obvious from Theorem~\ref{th:eqv-asym-expo} that such inequalities can be used also for 
uniform asymptotic stability.
The Lyapunov matrix in the inequalities will be revealed soon to be
$\xi$-dependent for ensuring the necessity of the condition.
If we restrict the Lyapunov matrix to a $\xi$-independent constant
matrix, the necessity does not hold in general.
Since such a special case of the condition is closely related with so
called quadratic stability, 
we also give some associated comments.
Then, we further derive another type of Lyapunov inequality condition that can be
used for stochastic systems having essentially bounded
coefficient matrices.

\subsection{Lyapunov inequalities for general stochastic systems}

To show the Lyapunov inequality characterizing exponential stability of 
system (\ref{eq:fr-sys}),
let us introduce the time shift operator
$S_k:({\bf R}^Z)^{{\bf Z}_+(k)}\rightarrow ({\bf R}^Z)^{{\bf N}_0}\
(k\in {\bf Z})$ for processes such that 
$\zeta^{0+}=S_k \xi^{k+}$
is defined by
	 $\zeta_0=\xi_{k},\ \zeta_1=\xi_{k+1},\ 
	 \ldots$;
note ${\bf N}_0={\bf Z}_+(0)$ by definition.
The introduction of this operator is merely a formality, and 
$\zeta_0$ for $\zeta^{0+}=S_k \xi^{k+}$ is ${\cal F}_k$-measurable for
each $k\in {\bf Z}$ since
it is nothing but $\xi_k$.
With this operator, we can show the following theorem giving a necessary
and sufficient inequality condition for exponential stability in the
most general case.

\begin{theorem}
\label{th:expo-lyap}
 Suppose the system (\ref{eq:fr-sys}) satisfies
 Assumption~\ref{as:bound}.
The following two conditions are equivalent.
\begin{enumerate}
 \item 
The system is exponentially stable in the second moment.
\item
There exist $\underline{\epsilon}_1, \overline{\epsilon}_1 \in {\bf R}_+$, $\lambda_1\in (0,1)$ and 
$P:({\bf R}^Z)^{{\bf N}_0}\rightarrow
	 {\bf S}^{n\times n}$ such that
\begin{align}
&
E_0[P(S_{k_0}\xi^{k_0+})]
\geq \underline{\epsilon}_1 I, \label{eq:P-lower}\\
&
E_0[P(S_{k_0}\xi^{k_0+})]
\leq \overline{\epsilon}_1 I, \label{eq:P-upper}\\
 & \hspace{-2mm}
E_0[\lambda_1^2 P(S_{k_0}\xi^{k_0+})-
A(\xi_{k_0})^T 
E_0[P(S_{k_0+1}\xi^{(k_0+1)+})| {\cal F}_{k_0}]\notag\\
&
\cdot A(\xi_{k_0})]\geq 0\ \ (\forall \widehat{\xi}^{(k_0-1)-}\in
 \widehat{\boldsymbol \Xi}_0; \forall k_0 \in {\bf Z}).
\label{eq:gen-lyap}
\end{align}
%
%
\end{enumerate}
\end{theorem}

\begin{IEEEproof}

2$\Rightarrow$1:
It follows from the inequality in (\ref{eq:gen-lyap}) that
\begin{align}
 &
E_0[x_{k_0+1}^T
E_0[P(S_{k_0+1}\xi^{(k_0+1)+})| {\cal F}_{k_0}]
 x_{k_0+1}]
\notag \\
\leq &
\lambda_1^2 x_{k_0}^T 
E_0[P(S_{k_0}\xi^{k_0+})]
x_{k_0}.
\end{align}
Since this inequality holds for
each $k_0 \in {\bf Z}$, every $\widehat{\xi}^{(k_0-1)-}\in
 \widehat{\boldsymbol \Xi}_0$ and every 
$x_{k_0}\in {\bf R}^n$, 
we have
\begin{align}
 &
E_0[x_{k+1}^T
E_0[P(S_{k+1}\xi^{(k+1)+})| {\cal F}_{k}]
 x_{k+1}| {\cal F}_{k-1}]
\notag \\
\leq &
\lambda_1^2 x_{k}^T 
E_0[P(S_k \xi^{k+})| {\cal F}_{k-1}]
x_{k}\ {\rm a.s.}\notag \\
&
(\forall k
 \in {\bf Z}_+(k_0); \forall x_{k_0}\in {\bf R}^n;
\forall \widehat{\xi}^{(k_0-1)-}\in
 \widehat{\boldsymbol \Xi}_0;
 \forall k_0 \in {\bf Z}),
\end{align}
which implies (by (\ref{eq:con-exp-2}))
\begin{align}
 &
E_0[x_{k+1}^T
E_0[P(S_{k+1} \xi^{(k+1)+})| {\cal F}_{k}]
 x_{k+1}]
\notag \\
\leq &
\lambda_1^2 
E_0[x_{k}^T 
E_0[P(S_k \xi^{k+})| {\cal F}_{k-1}]
x_{k}]\notag \\
&
(\forall k
 \in {\bf Z}_+(k_0); \forall x_{k_0}\in {\bf R}^n;
\forall \widehat{\xi}^{(k_0-1)-}\in
 \widehat{\boldsymbol \Xi}_0;
 \forall k_0 \in {\bf Z}).\label{eq:th3-pf-3}
\end{align}
A recursive use of this inequality
leads to
\begin{align}
&
E_0[x_{k}^T 
E_0[P(S_k \xi^{k+})| {\cal F}_{k-1}]
x_{k}] \notag \\
\leq
&
 \lambda_1^{2(k-k_0)}x_{k_0}^T 
E_0[P(S_{k_0} \xi^{k_0+})]
x_{k_0}\notag \\
&
(\forall k
 \in {\bf Z}_+(k_0); \forall x_{k_0}\in {\bf R}^n;
\forall \widehat{\xi}^{(k_0-1)-}\in
 \widehat{\boldsymbol \Xi}_0;
 \forall k_0 \in {\bf Z}).
\end{align}
For the left-hand side of this inequality, (\ref{eq:P-lower}) leads to
\begin{align}
&
\underline{\epsilon}_1 E_0[\|x_{k}\|^2] 
\leq
E_0[x_{k}^T 
E_0[P(S_k \xi^{k+})| {\cal F}_{k-1}]
x_{k}],
\end{align}
while for the right-hand side, (\ref{eq:P-upper}) leads to
\begin{align}
&
\lambda_1^{2(k-k_0)}x_{k_0}^T 
E_0[P(S_{k_0}\xi^{k_0+})]
x_{k_0}
 \leq
\overline{\epsilon}_1 \|x_{k_0}\|^2 \lambda_1^{2(k-k_0)}.
\end{align}
Hence, we have (\ref{eq:exp-def}) with $a=\overline{\epsilon}_1/\underline{\epsilon}_1$
and $\lambda=\lambda_1$,
which means by definition that
 the system is exponentially stable in the second moment.


\medskip
1$\Rightarrow$2:
For $\lambda\in (0,1)$ satisfying (\ref{eq:exp-def}),
take $\lambda_1$ such that
$\lambda<\lambda_1<1$ and define
\begin{align}
&
\Gamma_{k_2}(S_{k_1} \xi^{k_1+})
\notag \\
:=
&
\begin{cases}
I & (k_2=k_1)\\
(A(\xi_{k_2-1})/\lambda_1)\cdots 
(A(\xi_{k_1})/\lambda_1)
& (k_2\geq k_1+1)
\end{cases}\label{eq:def-Gamma}
\end{align}
for $k_1, k_2 \in {\bf Z}$ such that $k_2 \geq k_1$.
Then, (\ref{eq:exp-def}) can be rewritten as
\begin{align}
&
x_{k_0}^T E_0[\Gamma_k(S_{k_0}\xi^{k_0+})^T \Gamma_k(S_{k_0}\xi^{k_0+})]x_{k_0} 
\notag \\
\leq &x_{k_0}^T
(a(\lambda/\lambda_1)^{2(k-k_0)} I)x_{k_0}\notag \\ 
&
(\forall k
 \in {\bf Z}_+(k_0); \forall x_{k_0}\in {\bf R}^n;
\forall \widehat{\xi}^{(k_0-1)-}\in
 \widehat{\boldsymbol \Xi}_0;
 \forall k_0 \in {\bf Z}),
\label{eq:th2-pf-rw}
\end{align}
which implies 
\begin{align}
&
E_0[\Gamma_k(S_{k_0}\xi^{k_0+})^T \Gamma_k(S_{k_0}\xi^{k_0+})]
\leq 
a(\lambda/\lambda_1)^{2(k-k_0)} I\notag \\ 
&
(\forall k
 \in {\bf Z}_+(k_0); 
\forall \widehat{\xi}^{(k_0-1)-}\in
 \widehat{\boldsymbol \Xi}_0;
 \forall k_0 \in {\bf Z}).
\label{eq:th2-pf-rw-any}
\end{align}

We next define
\begin{align}
&
P_{K}(S_{k} \xi^{k+})
:=
\lambda_1^{-2}
\sum_{t=k}^{K}
\Gamma_{t}(S_{k}\xi^{k+})^T \Gamma_{t}(S_{k}\xi^{k+})
\label{eq:def-PKk}
\end{align}
for $K\geq k$.
Then, it satisfies
\begin{align}
&
\lambda_1^{2} 
P_{K}(S_{k_0} \xi^{k_0+})
- A(\xi_{k_0})^T
P_{K}(S_{k_0+1}\xi^{(k_0+1)+})
A(\xi_{k_0}) =I
\end{align}
for $K\geq k_0+1$,
and thus,
\begin{align}
&
 \lambda_1^{2} E_0[P_{K}(S_{k_0} \xi^{k_0+})]\notag \\
\geq&
E_0[A(\xi_{k_0})^T
 E_0[P_{K}(S_{k_0+1}\xi^{(k_0+1)+})| {\cal F}_{k_0}] A(\xi_{k_0})]\label{eq:K-finite}
\end{align}
by (\ref{eq:con-exp-2}) and (\ref{eq:con-exp-3}).
On the other hand,
(\ref{eq:def-PKk}) also implies that
the sequence of
\begin{align}
&
E_0[P_{K}(S_k\xi^{k+})]
=
\lambda_1^{-2}
\sum_{t={k}}^{K}
E_0[\Gamma_{t}(S_k \xi^{k+})^T \Gamma_{t}(S_k \xi^{k+})]
\end{align}
with respect to $K$ for each fixed $k$ is
monotonically non-decreasing under the
semi-order relation based on positive semidefiniteness,
i.e., 
\begin{align}
 &
E_0[P_{K}(S_k \xi^{k+})]\leq 
E_0[P_{K+1}(S_k \xi^{k+})].
\end{align}
In addition, 
it follows from
(\ref{eq:th2-pf-rw-any}) that
\begin{align}
&
E_0[P_{K}(S_k \xi^{k+})]
\leq
\lambda_1^{-2}
a
\left(\sum_{t=k_0}^{K}(\lambda/\lambda_1)^{2(t-k_0)}\right)I,
\label{eq:upper-proof}
\end{align}
whose right-hand side converges to a 
constant matrix as $K\rightarrow\infty$.
Hence, the conditional expectation of $P_\infty(S_k \xi^{k+})=P_{K}(S_k
 \xi^{k+})|_{K\rightarrow \infty}$ given 
${\xi}^{(k_0-1)-}=\widehat{\xi}^{(k_0-1)-}$
(i.e.,
$E_0[P_\infty(S_k \xi^{k+})]$) is bounded 
(for each $k_0\in {\bf Z}$, every $\widehat{\xi}^{(k_0-1)-}\in
 \widehat{\boldsymbol \Xi}_0$ and every $k \in {\bf Z}_+(k_0)$).
We take $P(\cdot)=P_{\infty}(\cdot)$, which itself is independent of $k_0$
 and $k$ because $A(\cdot)$ is.
By definition, this $P$
satisfies (\ref{eq:P-lower}) (by (\ref{eq:def-Gamma}) and (\ref{eq:def-PKk})) and
 (\ref{eq:P-upper}) (by (\ref{eq:upper-proof})) with
 appropriate $\underline{\epsilon}_1, \overline{\epsilon}_1
 \in {\bf R}_+$.
In addition, 
letting $K\rightarrow \infty$ in
(\ref{eq:K-finite}) leads to
(\ref{eq:gen-lyap}).
This completes the proof.
\end{IEEEproof}

The inequality (\ref{eq:gen-lyap}) is a Lyapunov
inequality for the system (\ref{eq:fr-sys}) satisfying
Assumption~\ref{as:bound}, which is a generalization of the usual Lyapunov
inequality for discrete-time linear deterministic systems.

Let $\lambda_{\rm min}$ and $\lambda_{1{\rm min}}$ be respectively the infimum of
$\lambda$ such that there exists $a\in {\bf R}_+$ satisfying
(\ref{eq:exp-def}) and that of $\lambda_1$ such that there exist 
$\underline{\epsilon}_1, \overline{\epsilon}_1 \in {\bf R}_+$ and 
$P:({\bf R}^Z)^{{\bf N}_0}\rightarrow
	 {\bf S}^{n\times n}$
satisfying (\ref{eq:P-lower})--(\ref{eq:gen-lyap}).
Then, since $\lambda_1$ in the part $1\Rightarrow 2$ of the above proof
can be taken arbitrarily close to $\lambda$, and since
$\lambda=\lambda_1$ in the proof of the opposite direction,
we have the following equality.
\begin{align}
&
\lambda_{\rm min} = \lambda_{1{\rm min}}
\end{align}
This implies that we can characterize the convergence rate of the
sequence $\left(\sqrt{E_0[\|x_k\|^2]}\right)_{k\in {\bf Z}_+(k_0)}$
by the inequality condition
(\ref{eq:P-lower})--(\ref{eq:gen-lyap}) without loss of generality.

However, if we are interested only in whether the system is stable and
not in the convergence rate, 
the Lyapunov
inequality without $\lambda_1$ shown in the following lemma would be sufficient.
\begin{lemma}
\label{lm:expo-lyap}
 Suppose the system (\ref{eq:fr-sys}) satisfies
 Assumption~\ref{as:bound}.
The following two conditions are equivalent.
\begin{enumerate}
\item
There exist $\underline{\epsilon}_1, \overline{\epsilon}_1 \in {\bf R}_+$, $\lambda_1\in (0,1)$ and 
$P:({\bf R}^Z)^{{\bf N}_0}\rightarrow
	 {\bf S}^{n\times n}$ satisfying (\ref{eq:P-lower})--(\ref{eq:gen-lyap}).
\item
There exist $\underline{\epsilon}_1, \overline{\epsilon}_1, \epsilon_1 \in {\bf R}_+$ and 
$P:({\bf R}^Z)^{{\bf N}_0}\rightarrow
	 {\bf S}^{n\times n}$ satisfying (\ref{eq:P-lower}),
	 (\ref{eq:P-upper}) and
\begin{align}
 & \hspace{-8mm}
E_0[P(S_{k_0}\xi^{k_0+})-
A(\xi_{k_0})^T 
E_0[P(S_{k_0+1}\xi^{(k_0+1)+})| {\cal F}_{k_0}]
A(\xi_{k_0})]\notag\\
&\geq \epsilon_1 I\ \ (\forall \widehat{\xi}^{(k_0-1)-}\in
 \widehat{\boldsymbol \Xi}_0; \forall k_0 \in {\bf Z}).
\label{eq:gen-lyap-nol}
\end{align}
\end{enumerate}
\end{lemma}

\begin{IEEEproof}
Adding $E_0[(1-\lambda_1^2)P(S_{k_0} \xi^{k_0+})]$ to (\ref{eq:gen-lyap}) and using
(\ref{eq:P-lower}) lead to (\ref{eq:gen-lyap-nol}) with
 $\epsilon_1=(1-\lambda_1^2)\underline{\epsilon}_1\ (>0)$.
The opposite assertion is obvious from (\ref{eq:P-upper}).
\end{IEEEproof}

As already stated, the inequality condition (\ref{eq:P-lower}),
(\ref{eq:P-upper}) and (\ref{eq:gen-lyap-nol}) (and thus,
(\ref{eq:P-lower})--(\ref{eq:gen-lyap})) is necessary and
sufficient not only for exponential stability but also for
uniform asymptotic stability 
(recall Theorem~\ref{th:eqv-asym-expo} and Fig.~\ref{fig:gen_relations})
under Assumption~\ref{as:bound}.

\subsection{Quadratic stability}

The Lyapunov inequalities (\ref{eq:gen-lyap}) and
(\ref{eq:gen-lyap-nol}) involve the $\xi$-dependent Lyapunov matrix, and
hence, the direct use of them for numerical analysis is considered not so
easy.
In the case of deterministic linear time-varying (or parameter-varying)
systems,
we sometimes consider restricting the Lyapunov matrix to a constant
matrix for ease of numerical analysis as in \cite{de1993h,xu2001quadratic}, and a similar idea might be useful for our
Lyapunov inequalities.
If we introduce the restriction $P(\cdot)=P_0$ for $P_0\in {\bf S}^{n\times n}$,
the inequality condition in Theorem~\ref{th:expo-lyap}
reduces to the following form:
there exist $\lambda_1\in (0,1)$ and $P_0\in {\bf S}^{n\times n}_+$ such
that
\begin{align}
&
E_0[\lambda_1^2 P_0-
A(\xi_{k_0})^T 
P_0 A(\xi_{k_0})]\geq 0\notag \\ 
&(\forall \widehat{\xi}^{(k_0-1)-}\in
 \widehat{\boldsymbol \Xi}_0; \forall k_0 \in {\bf Z}).
\label{eq:gen-lyap-const}
\end{align}
While this condition is only sufficient for exponential stability,
it is also necessary for quadratic stability defined in the following.
\begin{definition}[Quadratic Stability]
\label{df:quad-stab}
The system (\ref{eq:fr-sys}) is said to be quadratically stable
 if there exist $P_0\in {\bf S}^{n\times n}_+$ and $\lambda \in
 (0,1)$ such that
\begin{align}
&
E_0[x_{k+1}^T P_0 x_{k+1}]\leq \lambda^2 E_0[x_{k}^T P_0 x_{k}]\notag
\\
&
(\forall k \in {\bf Z}_+(k_0); \forall x_{k_0} \in {\bf R}^n; \forall \widehat{\xi}^{(k_0-1)-}\in \widehat{\boldsymbol \Xi}_0; \forall
k_0\in {\bf Z}).
\label{eq:quad-def}
\end{align}
\end{definition}

We can show that (\ref{eq:quad-def}) implies (\ref{eq:gen-lyap-const}) (i.e.,
the above necessity assertion) through taking $k=0$ and
$\lambda_1=\lambda$; the opposite direction is obvious from the proof of
Theorem~\ref{th:expo-lyap} (see the arguments around (\ref{eq:th3-pf-3})).

It is obvious from the introduced restriction that the equivalence
between exponential stability and quadratic stability does not hold in
general (see Fig.~\ref{fig:gen_relations}).
This is true even when Assumption~\ref{as:stationary} is additionally
satisfied.
However, in the case of i.i.d.\ processes \cite{Hosoe-TAC19}, 
the equivalence is known to hold.
We will revisit this special case as one of the selected applications later.

\subsection{Lyapunov inequalities under essential boundedness assumption}

In this section, we derived new Lyapunov inequalities
(\ref{eq:gen-lyap}) and (\ref{eq:gen-lyap-nol}) for
system (\ref{eq:fr-sys}) only with Assumption~\ref{as:bound},
which was a minimal requirement for defining the notions of 
second-moment stability.
In this subsection, we consider another assumption on system
(\ref{eq:fr-sys}) that is stronger than Assumption~\ref{as:bound} 
and derive different Lyapunov inequalities.
The assumption we use here is the following.
\begin{assumption}
\label{as:ess-bound}
There exists $M_3\in {\bf R}_+$ such that 
\begin{align}
&
|A_{ij}(\xi_{k_0})|<M_3\ {\rm a.s.}\notag \\ 
&
(\forall i, j = 1,\ldots,n;
\forall \widehat{\xi}^{(k_0-1)-}\in
 \widehat{\boldsymbol \Xi}_0; \forall k_0\in {\bf Z}),
\label{eq:as-ess-bound}
\end{align}%
where $|\cdot |$ denotes the absolute value.
\end{assumption}

By this assumption, the square entries of 
$A(\xi_{k})$ become essentially bounded.
Hence, $A(\xi_{k})$ satisfying this assumption
also satisfies Assumption~\ref{as:bound}.

With Assumption~\ref{as:ess-bound}, we can show the following theorem.
\begin{theorem}
\label{th:expo-lyap2}
 Suppose the system (\ref{eq:fr-sys}) satisfies
 Assumption~\ref{as:ess-bound}.
The following two conditions are equivalent.
\begin{enumerate}
\item
There exist $\underline{\epsilon}_1, \overline{\epsilon}_1 \in {\bf
	 R}_+$, $\lambda_1 \in (0,1)$ and 
$P:({\bf R}^Z)^{{\bf N}_0}\rightarrow
	 {\bf S}^{n\times n}$ satisfying (\ref{eq:P-lower})--(\ref{eq:gen-lyap}).
\item
There exist $\underline{\epsilon}_2, \overline{\epsilon}_2, \epsilon_2
	 \in {\bf R}_+$, $\lambda_2 \in (0,1)$ and 
$R:({\bf R}^Z)^{{\bf N}_0}\rightarrow
	 {\bf S}^{n\times n}$ such that
\begin{align}
&
E_0[R(S_{k_0}\xi^{k_0+})| {\cal F}_{k_0}]
\geq \underline{\epsilon}_2 I\ {\rm a.s.}, \label{eq:R-lower}\\
&
E_0[R(S_{k_0}\xi^{k_0+})| {\cal F}_{k_0}]
\leq \overline{\epsilon}_2 I\ {\rm a.s.}, \label{eq:R-upper}\\
 &
\lambda_2^2 
E_0[R(S_{k_0}\xi^{k_0+})| {\cal F}_{k_0}]\notag \\
&
-A(\xi_{k_0})^T
E_0[R(S_{k_0+1}\xi^{(k_0+1)+})| {\cal F}_{k_0}]
A(\xi_{k_0})\notag \\
&
 \geq \epsilon_2 I\ {\rm a.s.}\ \ (\forall \widehat{\xi}^{(k_0-1)-}\in
 \widehat{\boldsymbol \Xi}_0;
\forall k_0\in {\bf Z}).
\label{eq:gen-lyap2}
\end{align}
\end{enumerate}
\end{theorem}

The intuitive interpretation of the above theorem is that
$A(\xi_{k_0})$ in (\ref{eq:gen-lyap}) can be taken out from the conditional expectation as in
(\ref{eq:gen-lyap2}) through shifting time for conditions of conditional
expectations so that
$E_0[\cdot]$ is replaced by $E_0[\cdot | {\cal F}_{k_0}]$,
when the entries of $A(\xi_k)$ are essentially bounded (otherwise,
(\ref{eq:gen-lyap2}) does not make sense).
For example, uniformly distributed entries of $A(\xi_{k})$ can be dealt with in the
inequality condition (\ref{eq:R-lower})--(\ref{eq:gen-lyap2}) (while normally distributed entries cannot).
The proof of the theorem is as follows.
\begin{IEEEproof}
2$\Rightarrow$1:
Taking the conditional expectations $E_0[\cdot]$
for (\ref{eq:R-lower})--(\ref{eq:gen-lyap2})
leads to (\ref{eq:P-lower})--(\ref{eq:gen-lyap})
with $\underline{\epsilon}_1=\underline{\epsilon}_2$, 
$\overline{\epsilon}_1=\overline{\epsilon}_2$
and $P=R$.

\medskip
1$\Rightarrow$2:
It follows from (\ref{eq:gen-lyap}) that
\begin{align}
&
E_0[\lambda_1^2 P(S_{k_0+1}\xi^{(k_0+1)+})\notag \\
&-A(\xi_{k_0+1})^T 
E_0[P(S_{k_0+2}\xi^{(k_0+2)+})| {\cal F}_{k_0+1}]
A(\xi_{k_0+1})|
{\cal F}_{k_0}]\notag \\
&\geq 0
\ {\rm a.s.}\ (\forall \widehat{\xi}^{(k_0-1)-}\in
 \widehat{\boldsymbol \Xi}_0; \forall k_0\in {\bf Z}).
\label{eq:lyap-shift}
\end{align}
Take $\lambda_2$ satisfying $\lambda_1<\lambda_2<1$
and define
\begin{align}
&
\underline{\epsilon}_1^\prime:=(\lambda_2^{2}-\lambda_1^2) \underline{\epsilon}_1.
\end{align}
Then, (\ref{eq:lyap-shift}) together with (\ref{eq:P-lower}) leads us to
\begin{align}
&
E_0[\lambda_2^{2} P(S_{k_0+1}\xi^{(k_0+1)+})\notag\\
&
-A(\xi_{k_0+1})^T 
E_0[P(S_{k_0+2}\xi^{(k_0+2)+})| {\cal F}_{k_0+1}]
A(\xi_{k_0+1})|
{\cal F}_{k_0}]
\notag\\
&\geq (\lambda_2^{2}-\lambda_1^2) E_0[P(S_{k_0+1}\xi^{(k_0+1)+})| {\cal F}_{k_0}]\notag\\
&\geq \underline{\epsilon}_1^\prime I
\ {\rm a.s.}\ (\forall \widehat{\xi}^{(k_0-1)-}\in
 \widehat{\boldsymbol \Xi}_0; \forall k_0\in {\bf Z}). \label{eq:lyap-shift-tr}
\end{align}
Take
\begin{align}
R(S_k \xi^{k+})=A(\xi_k)^T 
P(S_{k+1} \xi^{(k+1)+})
A(\xi_k)+\underline{\epsilon}_1^\prime I.
\label{eq:take-R}
\end{align}
Then, since 
\begin{align}
&E_0[R(S_{k_0}\xi^{k_0+})|{\cal F}_{k_0}]
\notag \\
=&
A(\xi_{k_0})^T 
E_0[P(S_{k_0+1}\xi^{(k_0+1)+})|{\cal F}_{k_0}]
A(\xi_{k_0})+\underline{\epsilon}_1^\prime I
\label{eq:take-R}
\end{align}
holds,
(\ref{eq:P-lower}) leads to
 (\ref{eq:R-lower}) with $\underline{\epsilon}_2=\underline{\epsilon}_1^\prime>0$,
while (\ref{eq:P-upper}) and Assumption~\ref{as:ess-bound} lead to
(\ref{eq:R-upper}) with appropriate $\overline{\epsilon}_2$.
By using this $R$, 
(\ref{eq:lyap-shift-tr}) can be rewritten as
\begin{align}
&
\lambda_2^{2}E_0[P(S_{k_0+1}\xi^{(k_0+1)+})|
{\cal F}_{k_0}]
-E_0[R(S_{k_0+1}\xi^{(k_0+1)+})|
{\cal F}_{k_0}]\notag \\
&
\geq 0\ {\rm a.s.}\ 
(\forall \widehat{\xi}^{(k_0-1)-}\in
 \widehat{\boldsymbol \Xi}_0;
\forall k_0\in {\bf Z})
\end{align}
By post- and pre-multiplying $A(\xi_{k_0})$
 and its transpose respectively on this inequality, 
we further obtain
\begin{align}
&
\lambda_2^{2}E_0[A(\xi_{k_0})^T P(S_{k_0+1}\xi^{(k_0+1)+}) A(\xi_{k_0})|
{\cal F}_{k_0}]\notag \\
&
-A(\xi_{k_0})^T E_0[R(S_{k_0+1}\xi^{(k_0+1)+})|
{\cal F}_{k_0}]A(\xi_{k_0}) \geq 0\ {\rm a.s.}\notag \\
& (\forall \widehat{\xi}^{(k_0-1)-}\in
 \widehat{\boldsymbol \Xi}_0;
\forall k_0\in {\bf Z}),
\end{align}
which implies (\ref{eq:gen-lyap2}) with
 $\epsilon_2=\lambda_2^{2}\underline{\epsilon}_1^\prime\ (\in {\bf
 R}_+)$.
This completes the proof.
\end{IEEEproof}

Let $\lambda_{2{\rm min}}$ be the infimum of
$\lambda_2$ such that there exist 
$\underline{\epsilon}_2, \overline{\epsilon}_2, \epsilon_2 \in {\bf R}_+$ and 
$R:({\bf R}^Z)^{{\bf N}_0}\rightarrow
	 {\bf S}^{n\times n}$
satisfying (\ref{eq:R-lower})--(\ref{eq:gen-lyap2}).
Then, it follows from the above proof that $\lambda_{1{\rm min}} =
\lambda_{2{\rm min}}$, and thus,
\begin{align}
&
\lambda_{{\rm min}} = \lambda_{2{\rm min}}.
\end{align}

In addition, we can also show the following lemma.
\begin{lemma}
\label{lm:expo-lyap2}
 Suppose the system (\ref{eq:fr-sys}) satisfies
 Assumption~\ref{as:ess-bound}.
The following two conditions are equivalent.
\begin{enumerate}
\item
There exist $\underline{\epsilon}_2, \overline{\epsilon}_2, \epsilon_2
	 \in {\bf R}_+$, $\lambda_2 \in (0,1)$ and 
$R:({\bf R}^Z)^{{\bf N}_0}\rightarrow
	 {\bf S}^{n\times n}$ satisfying (\ref{eq:R-lower})--(\ref{eq:gen-lyap2}).
\item
There exist $\underline{\epsilon}_2, \overline{\epsilon}_2, \epsilon_2
	 \in {\bf R}_+$ and 
$R:({\bf R}^Z)^{{\bf N}_0}\rightarrow
	 {\bf S}^{n\times n}$ satisfying (\ref{eq:R-lower}),
	 (\ref{eq:R-upper}) and
\begin{align}
 &
E_0[R(S_{k_0}\xi^{k_0+})| {\cal F}_{k_0}]\notag \\
&-A(\xi_{k_0})^T
E_0[R(S_{k_0+1}\xi^{(k_0+1)+})| {\cal F}_{k_0}]
A(\xi_{k_0})\notag \\
& \geq \epsilon_2 I\ {\rm a.s.}\ \ (\forall \widehat{\xi}^{(k_0-1)-}\in
 \widehat{\boldsymbol \Xi}_0;
\forall k_0\in {\bf Z}).
\label{eq:gen-lyap2-nol}
\end{align}
\end{enumerate}
\end{lemma}

Lyapunov inequalities (\ref{eq:gen-lyap}) and
(\ref{eq:gen-lyap-nol}) correspond to a generalization of the earlier result in \cite{Hosoe-TAC19} for
systems with dynamics determined by an i.i.d.\ process while another
type of Lyapunov inequalities (\ref{eq:gen-lyap2}) and (\ref{eq:gen-lyap2-nol})
correspond to that in \cite{Costa-TAC14} for 
systems with dynamics determined by a Markov process, both of which will
be revisited in the following section.
As already stated, the results in the two earlier frameworks do not cover each
other, and we had a question of what is the essential reason for this
difference.
According to Theorem~\ref{th:expo-lyap2}, the two types of Lyapunov
inequalities can be shown to be equivalent under
Assumption~\ref{as:ess-bound}.
This implies that the assumptions on the process $\xi$ to be i.i.d.\ or
Markov are not essential, and only it is important in 
the selection of the type of Lyapunov inequalities
whether Assumption~\ref{as:ess-bound} is
satisfied or not 
 (note a Lyapunov inequality of the same type as (\ref{eq:gen-lyap}) and
(\ref{eq:gen-lyap-nol}) can be always derived under
Assumption~\ref{as:bound}).
This is the answer to the question, which was led to by our
new unified framework for second-moment stability of systems with
stochastic dynamics.

Since Assumption~\ref{as:ess-bound} is stronger than 
Assumption~\ref{as:bound}, the relations in
Fig.~\ref{fig:gen_relations} automatically hold even under Assumption~\ref{as:ess-bound}. 
In addition, the above arguments in turn imply that 
we can derive a Lyapunov inequality in the type of (\ref{eq:gen-lyap2})
and (\ref{eq:gen-lyap2-nol}) even for the i.i.d.\ case under
Assumption~\ref{as:ess-bound}, and that in the type of 
(\ref{eq:gen-lyap}) and
(\ref{eq:gen-lyap-nol}) 
even for the Markov case under Assumption~\ref{as:bound}.
The associated results will be also shown in the following section.

\section{Selected Applications}
\label{sc:app}

In this section, we demonstrate usefulness of the results obtained in the
preceding sections through providing their applications.
Our results, together with additional assumptions on $\xi$,
readily lead us to Lyapunov inequalities for the corresponding special
cases of systems.
We here deal with temporally-independent processes, Markov processes and
polytopic martingales for determining system dynamics as selected examples, and discuss associated stability
conditions.
These arguments not only reveal the connections between earlier and our
results but also lead to generalizations of the former.

\subsection{Temporally-Independent Process Case}

We first consider the following assumption on $\xi$.
\begin{assumption}
\label{as:independent}
For $\xi=(\xi_k)_{k\in {\bf Z}}$,
the random vectors $\xi_{k}\ (k\in {\bf Z})$ are independently distributed.
\end{assumption}

We call the process $\xi$ satisfying this assumption a temporally-independent process.
With this assumption,
the Lyapunov matrix in (\ref{eq:P-lower})--(\ref{eq:gen-lyap}) becomes
independent of $\widehat{\xi}^{(k_0-1)-}$ for each $k_0$.
Hence, the associated conditional expectations 
can be replaced by the standard expectations.
Let $P_{k_0}=E[P(S_{k_0}\xi^{k_0+})]$.
Then, this implies
that (\ref{eq:P-lower})--(\ref{eq:gen-lyap})
reduce to
\begin{align}
&
P_{k_0}
\geq \underline{\epsilon}_1 I \label{eq:ind-P-lower}, \\
&
P_{k_0}
\leq \overline{\epsilon}_1 I,\label{eq:ind-P-upper}\\
 & 
E[\lambda_1^2 P_{k_0}-
A(\xi_{k_0})^T
P_{k_0+1} 
A(\xi_{k_0})]\geq 0\ \ (\forall k_0 \in {\bf Z})\label{eq:ind-lyap}
\end{align}
under Assumption~\ref{as:independent}, respectively.
In this manner, Assumption~\ref{as:independent} can lead us to Lyapunov
inequalities involving a deterministic (time-varying) Lyapunov matrix.

We further consider the situation where Assumptions~\ref{as:stationary} and
\ref{as:independent} are both satisfied.
This corresponds to the assumption that $\xi$ is an i.i.d.\ process,
which is used in \cite{Hosoe-TAC19}.
Then, the above Lyapunov matrix $P_{k_0}$ becomes time-invariant, since
\begin{align}
&
E[P(S_{k_0}\xi^{k_0+})]=E[P(S_{k_0+1}\xi^{(k_0+1)+})]
\end{align}
for each $k_0\in {\bf Z}$.
This implies that the corresponding 
Lyapunov inequality can be described with
a deterministic time-invariant Lyapunov matrix.
In particular, for $P_0 \in {\bf S}^{n\times n}_+$,
\begin{align}
&
E[\lambda_1^2 P_0-
A(\xi_{k_0})^T 
P_0
A(\xi_{k_0})]\geq 0\ \ (\forall k_0 \in {\bf Z})\label{eq:ind-lyap-phomo}
\end{align}
if and only if
\begin{align}
&
E[\lambda_1^2 P_0-
A(\xi_{0})^T 
P_0
A(\xi_{0})]\geq 0\label{eq:homo-ind-lyap}
\end{align}
in this situation.
Hence, the following corollary holds.
\begin{corollary}
\label{cr:homo-ind-lyap}
 Suppose the system (\ref{eq:fr-sys}) satisfies
 Assumptions~\ref{as:bound}, \ref{as:stationary} and \ref{as:independent}.
The system is exponentially stable in the second moment if and only if
there exist $\lambda_1\in (0,1)$ and 
$P_0\in {\bf S}_+^{n\times n}$ satisfying (\ref{eq:homo-ind-lyap}).
\end{corollary}

The Lyapunov inequality (\ref{eq:homo-ind-lyap}) is nothing but that in
\cite{Hosoe-TAC19} (see also the dotted arrow in Fig.~\ref{fig:gen_relations}).
As was shown above, the Lyapunov inequalities derived in the present
paper for systems with
general stochastic dynamics can easily lead us to those for special
cases of systems (the Lyapunov inequality (\ref{eq:ind-lyap}) can be seen
as an extension of the earlier result (\ref{eq:homo-ind-lyap})).

Essentially the same techniques can be applied to 
(\ref{eq:R-lower})--(\ref{eq:gen-lyap2}).
However, since $R(S_{k_0}\xi^{k_0+})$ is not independent of $\xi_{k_0}$
even under Assumption~\ref{as:independent},
the conditional expectation 
$E_0[R(S_{k_0}\xi^{k_0+})| {\cal F}_{k_0}]$
cannot be
replaced by the standard expectation; it becomes a
random matrix depending on $\xi_{k_0}$.
We denote such a random matrix by $R_{k_0}(\xi_{k_0})$.
Then, taking account of this difference and
Theorems~\ref{th:expo-lyap} and \ref{th:expo-lyap2}
lead us to the following corollary.
\begin{corollary}
\label{cr:homo-ind-lyap2}
 Suppose the system (\ref{eq:fr-sys}) satisfies
 Assumptions~\ref{as:stationary}, \ref{as:ess-bound} and \ref{as:independent}.
The system is exponentially stable in the second moment if and only if
there exist $\underline{\epsilon}_2, \overline{\epsilon}_2, \epsilon_2
	 \in {\bf R}_+$, $\lambda_2 \in (0,1)$ and 
$R_0:{\bf R}^Z\rightarrow
	 {\bf S}^{n\times n}$ such that
\begin{align}
&
R_{0}(\xi_0)
\geq \underline{\epsilon}_2 I\ {\rm a.s.},\\
&
R_{0}(\xi_0)
\leq \overline{\epsilon}_2 I\ {\rm a.s.},\\
 &
\lambda_2^2 
R_0(\xi_0)
-A(\xi_{0})^T
E[R_0(\xi_0)]
A(\xi_{0}) \geq \epsilon_2 I\ \ {\rm a.s.}
\end{align}
\end{corollary}

If we introduce the support ${\boldsymbol \Xi}_0$ of $\xi_0$,
the inequality condition in this corollary can be rewritten as
\begin{align}
&
R_0(\xi_\star)
\geq \underline{\epsilon}_2 I, \label{eq:homo-ind-R-lower2}\\
&
R_0(\xi_\star)
\leq \overline{\epsilon}_2 I, \label{eq:homo-ind-R-upper2}\\
 &
\lambda_2^2 
R_0(\xi_\star)
-A(\xi_{\star})^T
E[R_0(\xi_0)]
A(\xi_{\star}) \geq \epsilon_2 I\ \ (\forall \xi_\star \in {\boldsymbol \Xi}_0),
\end{align}
which might be easier to interpret.

Compared to Corollary~\ref{cr:homo-ind-lyap}, the inequality condition
in Corollary~\ref{cr:homo-ind-lyap2} does not require us to deal with the
expectation of square entries of $A(\xi_0)$, which
might enable us to extend the inequality conditions toward
other control problems by using linear matrix inequality (LMI)
optimization techniques \cite{Boyd-book,Svariable-Ebihara-book} more
easily.
Instead, however, we have to directly deal with a random matrix
(i.e., $R_0(\xi_0)$) as a
decision variable in Corollary~\ref{cr:homo-ind-lyap2}, which
deteriorates the tractability of the condition in numerical analysis.
Since the inequality condition in Corollary~\ref{cr:homo-ind-lyap} can
also be extended (at least) to that for stabilization synthesis \cite{Hosoe-TAC19,Hosoe-LCSS19}, and since Assumption~\ref{as:ess-bound} is stronger
than Assumption~\ref{as:bound},
the superiority of Corollary~\ref{cr:homo-ind-lyap2}
over Corollary~\ref{cr:homo-ind-lyap} might be limited at this moment.

In the above, we discussed the case of stationary processes.
The associated assumption (i.e., Assumption~\ref{as:stationary}), however, can be
easily alleviated in our framework; this is obvious because we already
obtained (\ref{eq:ind-lyap}) without Assumption~\ref{as:stationary}.
For example, let us consider the following assumption about 
$N$-periodically stationary processes.
\begin{assumption}
\label{as:periodic-homo}
The stochastic process $\xi$ 
is $N$-periodically stationary; i.e., 
for each $i=0,1,\ldots,N-1$,
none of the characteristics of $\xi_{\kappa N + i}$ changes with
 $\kappa\in {\bf Z}$.
\end{assumption}

Then, the following periodic version of Corollary~\ref{cr:homo-ind-lyap} can be
obtained from Theorem~\ref{th:expo-lyap} (that of Corollary~\ref{cr:homo-ind-lyap2} is omitted but
easily obtained from Theorem~\ref{th:expo-lyap2}).
\begin{corollary}
\label{cr:phomo-ind-lyap}
 Suppose the system (\ref{eq:fr-sys}) satisfies
 Assumptions~\ref{as:bound}, \ref{as:independent} and \ref{as:periodic-homo}.
The system is exponentially stable in the second moment if and only if
there exist $\lambda_1\in (0,1)$ and 
$P_k\in {\bf S}_+^{n\times n}\ (k=0,\ldots,N-1)$ such that
\begin{align}
&
E[\lambda_1^2 P_k-
A(\xi_{k})^T 
P_{k+1}
A(\xi_{k})] \geq 0\ \ (k=0,\ldots,N-1),
\end{align}
where $P_{N}=P_{0}$.
\end{corollary}

The key in this corollary is that $P_{k_0}$ in
(\ref{eq:ind-P-lower})--(\ref{eq:ind-lyap}) becomes $N$-periodic under Assumption~\ref{as:periodic-homo}.
It is obvious that Corollary~\ref{cr:homo-ind-lyap} is a special case of
Corollary~\ref{cr:phomo-ind-lyap} under $N=1$.


\subsection{Markov Process Case}

We next consider the following assumption on $\xi$ about Markov processes.
\begin{assumption}
\label{as:markov}
The stochastic process $\xi$ 
has the Markov property;
i.e., for each subset ${\boldsymbol \Xi} \subset {\bf R}^Z$ and every
$i, j\in {\bf Z}$ such that $i<j$,
\begin{align}
{\rm Pr}(\xi_j\in {\boldsymbol \Xi}  |\, \xi_i, \xi_{i-1}, \ldots)={\rm Pr}(\xi_j\in
 {\boldsymbol \Xi} |\, \xi_i),
\end{align}
where ${\rm Pr}(\cdot | \cdot)$ denotes the conditional probability.
\end{assumption}

Under this assumption, the conditional expectation $E_0$ can be
simplified as
\begin{align}
&
E_0[\cdot]=E[\cdot | \xi_{k_0-1}=\xi_\star],\\
&
E_0[\cdot| {\cal F}_k]=E[\cdot | \xi_{k}]\ \ (k\geq k_0)
\end{align}
for $\xi_\star$ belonging to the support of $\xi_{k_0-1}$ denoted by ${\boldsymbol \Xi}_{k_0-1}$.
Hence, $E_0[P(S_{k_0}\xi^{k_0+})]$ and 
$E_0[P(S_{k_0+1}\xi^{(k_0+1)+})| \xi_{k_0}]$ in
(\ref{eq:P-lower})--(\ref{eq:gen-lyap}) can be respectively
simplified as $P_{k_0}(\xi_\star)$ and $P_{k_0+1}(\xi_{k_0})$ with an appropriate
time-varying function $P_{k_0}: {\bf R}^Z \rightarrow {\bf S}^{n\times n}$.
That is, (\ref{eq:P-lower})--(\ref{eq:gen-lyap}) 
reduce to
\begin{align}
&
P_{k_0}(\xi_\star) \geq \underline{\epsilon}_1 I \label{eq:mar-P-lower}, \\
&
P_{k_0}(\xi_\star)
\leq \overline{\epsilon}_1 I,\label{eq:mar-P-upper}\\
&
E[\lambda_1^2 P_{k_0}(\xi_\star)-
A(\xi_{k_0})^T 
P_{k_0+1}(\xi_{k_0})
A(\xi_{k_0})|\, \xi_{k_0-1}= \xi_\star]\geq 0\notag\\
&
 (\forall \xi_\star \in {\boldsymbol \Xi}_{k_0-1}; \forall k_0 \in {\bf Z}).\label{eq:mar-lyap}
\end{align}
This, together with Assumption~\ref{as:stationary} about stationary $\xi$ and
Theorem~\ref{th:expo-lyap},
leads us to the following corollary.
\begin{corollary}
\label{cr:homo-mar-lyap}
 Suppose the system (\ref{eq:fr-sys}) satisfies
 Assumptions~\ref{as:bound}, \ref{as:stationary} and \ref{as:markov}.
The system is exponentially stable in the second moment if and only if
there exist $\lambda_1\in (0,1)$ and a time-invariant function
$P_0: {\bf R}^Z \rightarrow {\bf S}_+^{n\times n}$ such that
\begin{align}
&
P_{0}(\xi_\star) \geq \underline{\epsilon}_1 I \label{eq:homo-mar-P-lower}, \\
&
P_{0}(\xi_\star)
\leq \overline{\epsilon}_1 I,\label{eq:homo-mar-P-upper}\\
&
E[\lambda_1^2 P_0(\xi_{\star})-
A(\xi_{0})^T 
P_0(\xi_{0})
A(\xi_{0})|\, \xi_{-1}= \xi_\star]\geq 0\notag \\
&
(\forall \xi_\star\in {\boldsymbol \Xi}_{-1}).\label{eq:homo-mar-lyap}
\end{align}
\end{corollary}

In a similar fashion, Theorem~\ref{th:expo-lyap2} also leads us to the
following corollary.
%
%
\begin{corollary}
\label{cr:homo-mar-lyap2}
 Suppose the system (\ref{eq:fr-sys}) satisfies
 Assumptions~\ref{as:stationary}, \ref{as:ess-bound} and \ref{as:markov}.
The system is exponentially stable in the second moment if and only if
there exist $\underline{\epsilon}_2, \overline{\epsilon}_2, \epsilon_2
	 \in {\bf R}_+$, $\lambda_2 \in (0,1)$ and 
$R_0:{\bf R}^Z \rightarrow {\bf S}_+^{n\times n}$ such that
 (\ref{eq:homo-ind-R-lower2}), (\ref{eq:homo-ind-R-upper2}) and
\begin{align}
&
\lambda_2^2 
R_0(\xi_\star)
-A(\xi_{\star})^T
E[R_0(\xi_{0})| \xi_{-1}=\xi_\star]
A(\xi_{\star})\geq \epsilon_2 I\notag \\
&
(\forall \xi_\star\in {\boldsymbol \Xi}_{-1}).
\end{align}
\end{corollary}

The Lyapunov inequality in Corollary~\ref{cr:homo-mar-lyap2} is essentially the same as that in \cite{Costa-TAC14}.
In contrast to Corollary~\ref{cr:homo-ind-lyap} (i.e.,
the temporally-independent process case), 
the Lyapunov matrix in Corollary~\ref{cr:homo-mar-lyap} does not become
a constant matrix even when Assumption~\ref{as:stationary} is used.
Hence, the tractability of the condition in
Corollary~\ref{cr:homo-mar-lyap} in numerical analysis 
would not be much different from 
that of the condition in Corollary~\ref{cr:homo-mar-lyap2} in the
Markov process case.
The difference between the two conditions is related with whether
$A(\xi_k)$ is essentially bounded or not (i.e., with or without
Assumption~\ref{as:ess-bound}).
In other words,
$A(\xi_0)$ should be contained in the (conditional) expectation as in
(\ref{eq:homo-mar-lyap}) of Corollary~\ref{cr:homo-mar-lyap}
unless Assumption~\ref{as:ess-bound} is satisfied.

As is the case with temporally-independent processes,
Theorems~\ref{th:expo-lyap} and \ref{th:expo-lyap2} further lead us to
the following periodic versions of Corollaries~\ref{cr:homo-mar-lyap}
and \ref{cr:homo-mar-lyap2}, which themselves would be new (a similar
comment also applies to the other cases of application).

\begin{corollary}
\label{cr:phomo-mar-lyap}
 Suppose the system (\ref{eq:fr-sys}) satisfies
 Assumptions~\ref{as:bound}, \ref{as:periodic-homo} and \ref{as:markov}.
The system is exponentially stable in the second moment if and only if
there exist $\lambda_1\in (0,1)$ and 
$P_k: {\bf R}^Z \rightarrow {\bf S}_+^{n\times n}\ (k=0,\ldots,N-1)$ such that
\begin{align}
&
P_k(\xi_{\star})
\geq \underline{\epsilon}_1 I, \label{eq:phomo-mar-P-lower}\\
&
P_k(\xi_{\star})
\leq \overline{\epsilon}_1 I, \label{eq:phomo-mar-P-upper}\\
 &
E[\lambda_1^2 P_k(\xi_{\star})-
A(\xi_{k})^T 
P_{k+1}(\xi_{k})
A(\xi_{k})|
\xi_{k-1}=\xi_\star]\notag\geq 0\noindent \\
&
(\forall \xi_\star\in {\boldsymbol \Xi}_{k-1}; k=0,\ldots,N-1),\label{eq:phomo-mar-lyap}
\end{align}
where $P_{N}=P_0$.
\end{corollary}

\begin{corollary}
\label{cr:phomo-mar-lyap2}
 Suppose the system (\ref{eq:fr-sys}) satisfies
 Assumptions~\ref{as:ess-bound}, \ref{as:periodic-homo} and \ref{as:markov}.
The system is exponentially stable in the second moment if and only if
there exist $\underline{\epsilon}_2, \overline{\epsilon}_2, \epsilon_2
	 \in {\bf R}_+$, $\lambda_2 \in (0,1)$ and 
$R_k:{\bf R}^Z \rightarrow {\bf S}_+^{n\times n}\ (k=0,\ldots,N-1)$ such that
\begin{align}
&
R_k(\xi_{\star})
\geq \underline{\epsilon}_2 I, \\
&
R_k(\xi_{\star})
\leq \overline{\epsilon}_2 I, \\
&
\lambda_2^2 
R_k(\xi_{\star})
-A(\xi_{\star})^T
E[R_{k+1}(\xi_{k})| \xi_{k-1}=\xi_\star]
A(\xi_{\star})\geq \epsilon_2 I\notag\\
&(\forall \xi_\star\in {\boldsymbol \Xi}_{k-1}; k=0,\ldots,N-1),
\end{align}
where $R_{N}=R_0$.
\end{corollary}

\subsection{Polytopic Martingale Case}

The preceding two subsections exemplified the generality of our
results through clarifying their connections with some earlier
studies.
In particular, the arguments implied that the Lyapunov inequalities
individually derived in the earlier studies can be seen as special cases of
our results, which were led to in this section just by introducing associated assumptions.
In this subsection, we further show potentials of our
results by introducing an assumption uncommon in the sense that the
corresponding stochastic systems have not been dealt with as the target
for analysis and synthesis in the field of control, to the best
knowledge of the authors.
Interestingly, the associated results will also provide us with a
new insight into a well-known robust stability condition for uncertain
deterministic systems.

The following is the assumption we use in this subsection.
\begin{assumption}
\label{as:martingale}
For each $k_0\in {\bf Z}$ and every $\widehat{\xi}^{(k_0-1)-}\in \widehat{\boldsymbol \Xi}_0$,
the stochastic process $\xi$ 
satisfies the following conditions.
\begin{enumerate}
\item[1a)] For each $k\in {\bf Z}_+(k_0)$, $\xi_k$ is ${\cal F}_k$-measurable
	   (this is automatically satisfied by the present definition of
	   ${\cal F}_k$).
\item[1b)] For each $k\in {\bf Z}_+(k_0)$, $E_0[\|\xi_k\|]<\infty$ (this
		 is automatically satisfied by the following condition~2).
\item[1c)] For each $k\in {\bf Z}_+(k_0)$, 
\begin{align}
 & E_0[\xi_{k+1}| {\cal F}_k]=\xi_{k}\ {\rm a.s.}
\end{align}
\item[2)] The support of $\xi_{k_0}$ belongs to (or given by)
\begin{align}
&
 {\bf E}^Z
:=
\left\{\theta \in {\bf R}^Z\Bigg| \theta_i\geq 0\,
 (i=1,\ldots,Z), \sum_{i=1}^Z \theta_i=1\right\},
\end{align}
where $\theta_i$ is the $i$th entry of $\theta$.
\end{enumerate}
\end{assumption}

Condition~1 in this assumption corresponds to the definition of
martingales \cite{Klenke-book} (so $\xi$ satisfying this assumption is a martingale).
In addition to it, we also consider condition~2 for restricting
$A(\xi_k)$ in (\ref{eq:fr-sys}) to
a polytopically-random matrix later.
Obviously, $\xi$ becomes neither temporally-independent
nor Markovian only with this assumption.
Hence, the system (\ref{eq:fr-sys}) with such $\xi$ cannot be dealt with
in the frameworks of the earlier studies referred to in the preceding subsections.

For given deterministic constant matrices $A^{(i)}\
(i=1,\ldots,Z)$, let system (\ref{eq:fr-sys}) further satisfy the
following assumption.
\begin{assumption}
\label{as:polytope}
The function $A: {\bf E}^Z \rightarrow {\bf R}^{n\times n}$ is given by
\begin{align}
&
A(\theta)=\sum_{i=1}^{Z}\theta_{i} A^{(i)} \label{eq:A-poly}
\end{align}
for $A^{(i)}\in {\bf R}^{n\times n}$ and $\theta\in {\bf E}^Z$.
\end{assumption}

This assumption, together with Assumption~\ref{as:martingale}, implies
that $A(\xi_k)$ takes values only in the polytope defined with
the vertices $A^{(i)}\ (i=1,\ldots,Z)$; in particular, the sequence of
such $A(\xi_k)$ with respect to $k$ is a martingale because
\begin{align}
E[A(\xi_{k+1})| {\cal F}_k]=\sum_{i=1}^{Z}E[\xi_{i(k+1)}| {\cal F}_k] A^{(i)}=A(\xi_k)\ {\rm a.s.}
\end{align}
for $\xi_k=[\xi_{1k},\ldots,\xi_{Zk}]^T$.

Since Assumptions~\ref{as:martingale} and \ref{as:polytope} let Assumption~\ref{as:ess-bound}
be automatically satisfied, we consider deriving a stability condition
 based on Theorem~\ref{th:expo-lyap2} rather than Theorem~\ref{th:expo-lyap}.
To this end, 
let us consider the Lyapunov matrix given by
\begin{align}
R(S_k\xi^{k+})=R_0(\xi_k)=\sum_{i=1}^{Z}\xi_{ik} R_0^{(i)}\label{eq:R-poly}
\end{align}
for $R_0^{(i)}\in {\bf S}_+^{n\times n}\ (i=1,\ldots,Z)$,
associated with Assumption~\ref{as:polytope}.
Because of this restriction, the corresponding
 stability condition becomes conservative.
However, it has the advantage that $(R_0(\xi_k))_{k\in {\bf Z}}$ also
 becomes a martingale, i.e.,
\begin{align}
 &
E[R_0(\xi_{k+1})| {\cal F}_k]=R_0(\xi_k)\ {\rm a.s.}, \label{eq:martin-R}
\end{align}
which will be a key in driving a tractable stability condition
with Theorem~\ref{th:expo-lyap2}.

With Assumptions~\ref{as:martingale} and \ref{as:polytope} and (\ref{eq:R-poly}),
the inequality condition (\ref{eq:gen-lyap2}) in Theorem~\ref{th:expo-lyap2} reduces to
\begin{align}
&
 \lambda_2^2 
R_0(\xi_{k_0})
-A(\xi_{k_0})^T
E[R_0(\xi_{k_0+1})| {\cal F}_{k_0}]
A(\xi_{k_0})\geq \epsilon_2 I\ {\rm a.s.}\notag\\
&(\forall \widehat{\xi}^{(k_0-1)-}\in \widehat{\boldsymbol \Xi}_0; \forall k_0 \in {\bf Z}).
\label{eq:martin-pre}
\end{align}
(note $E[R_0(\xi_{k_0})| {\cal F}_{k_0}]=R_0(\xi_{k_0})$); the other
inequality
conditions in the theorem are automatically
satisfied under Assumption~\ref{as:martingale} and (\ref{eq:R-poly}).
Then, (\ref{eq:martin-pre}) further reduces to 
\begin{align}
&
 \lambda_2^2 
R_0(\xi_{k_0})
-A(\xi_{k_0})^T
R_0(\xi_{k_0})
A(\xi_{k_0})\geq \epsilon_2 I\ {\rm a.s.}\notag\\
& 
(\forall \widehat{\xi}^{(k_0-1)-}\in \widehat{\boldsymbol \Xi}_0; \forall k_0 \in {\bf Z})
\label{eq:martin-pre2}
\end{align}
by (\ref{eq:martin-R}).
Hence, by using the $S$-variable (i.e., auxiliary variable) technique \cite{Svariable-Ebihara-book},
we can show that
there exists $\epsilon_2\in {\bf R}_+$ satisfying
(\ref{eq:martin-pre2}) (i.e., (\ref{eq:gen-lyap2})) if
there exist $\epsilon_2^\prime\in {\bf R}_+$ and
$S\in {\bf R}^{2n\times n}$ such that
\begin{align}
&
\begin{bmatrix}
 \lambda_2^2 
R_0(\xi_{k_0}) & 0\\
0 & -R_0(\xi_{k_0})
\end{bmatrix}
+
{\rm He}\left(S \begin{bmatrix}
				 A(\xi_{k_0}) & I
				\end{bmatrix}\right)
\geq \epsilon_2^\prime I\ {\rm a.s.}\notag\\
& (\forall
 \widehat{\xi}^{(k_0-1)-}\in \widehat{\boldsymbol \Xi}_0; \forall k_0 \in {\bf Z}),
\end{align}
%
where ${\rm He}(\cdot):=(\cdot)+(\cdot)^T$ for the square matrix $(\cdot)$.
Noting that 
the present $A$ and $R_0$
have the polytopic structures (\ref{eq:A-poly}) and (\ref{eq:R-poly}),
this immediately leads us to the following theorem.
\begin{theorem}
 \label{th:poly-martingale}
Suppose the system (\ref{eq:fr-sys}) satisfies
 Assumptions~\ref{as:martingale} and \ref{as:polytope}.
The system is exponentially stable in the second moment if
there exist $\lambda_2 \in (0, 1)$, $R_0^{(i)}\in {\bf S}_+^{n\times n}\
 (i=1,\ldots,Z)$ and $S\in {\bf R}^{2n\times n}$ such that
\begin{align}
&
\begin{bmatrix}
 \lambda_2^2 
R_0^{(i)} & 0\\
0 & -R_0^{(i)}
\end{bmatrix}
+
{\rm He}\left(S\begin{bmatrix}
			A^{(i)} & I	
			   \end{bmatrix}\right)
>0\ \ (i=1,\ldots,Z).
\label{eq:poly-martingale-expo}
\end{align}
%
%
\end{theorem}


Let $\widetilde{\bf \Xi}$ denote the set of processes $\xi$ satisfying
Assumption~\ref{as:martingale}.
Then, the above theorem actually implies that the system is stable for each
$\xi\in \widetilde{\bf \Xi}$ if a solution of (\ref{eq:poly-martingale-expo})
exists.
Hence, the above theorem gives a robust stability condition of the
system with respect to $\xi\in \widetilde{\bf \Xi}$.

Here, recall that $\xi$ given with $\xi_k=\theta\ (\forall k \in
{\bf Z})$ for the
deterministic vector $\theta\in{\bf E}^Z$ (i.e., the deterministic
process taking only $\theta$) satisfies
Assumption~\ref{as:martingale}, and hence, belongs to $\widetilde{\bf \Xi}$.
This special case corresponds to nothing but the deterministic system with the
polytopic uncertain parameter $\theta\in{\bf E}^Z$.
The inequality condition (\ref{eq:poly-martingale-expo}) in
Theorem~\ref{th:poly-martingale} is conventionally used as a sufficient
condition for robust stability of such a special case of systems.
However, according to our theorem, the inequality condition actually
ensures robust stability of the systems not only with such deterministic
time-invariant $\xi$ but also with stochastic time-varying $\xi$ satisfying
Assumption~\ref{as:martingale}.
This would not been known in the field of robust control, and in turn
demonstrates potentials of the results in this paper.
Since the form of the inequality condition in
(\ref{eq:poly-martingale-expo}) is consistent with that for
deterministic uncertain systems, it can be readily extended, e.g., toward
synthesis of robustly stabilizing state feedback, as is the case with deterministic systems.

\begin{remark}
In Theorem~\ref{th:poly-martingale}, if we confine $S$ to $[0, G^T]^T$
for $G\in {\bf R}^{n\times n}$, then the inequality
(\ref{eq:poly-martingale-expo}) reduces to
\begin{align}
&
\begin{bmatrix}
 \lambda_2^2 
R_0^{(i)} & A^{(i)T}G^T\\
GA^{(i)} & G+G^T-R_0^{(i)}
\end{bmatrix}
>0\ \ (i=1,\ldots,Z),
\label{eq:poly-martingale-expo-sp}
\end{align}
which is essentially the same as the inequality condition in Theorem~2
in \cite{extended-Oliveira-SCL99} for uncertain 
deterministic time-invariant systems.
One might be more familiar with this special case, to which a
 comment similar to (\ref{eq:poly-martingale-expo}) applies; that is,
it follows from our results that the inequality condition ensures
 robust stability of the systems not only for
 deterministic polytopic uncertainties but also for polytopic martingale
 uncertainties.
\end{remark}

\begin{remark}
In the case of deterministic systems, parameter-dependent Lyapunov
 inequalities for deterministic time-varying 
 uncertainties have been also studied, e.g., in \cite{Daafouz-SCL01}, and one might be also
 interested
 in the relationship of Theorem~\ref{th:poly-martingale} with this approach.
Since $A(\xi_k)$ satisfying Assumptions~\ref{as:martingale} and
 \ref{as:polytope} takes a value only in a ($k$-independent) polytope at each $k$,
robust stability for that polytope in the deterministic sense leads us to
 robust second-moment stability, which is dealt with in
 Theorem~\ref{th:poly-martingale}; deterministic stability is stronger
 than stochastic stability, in general.
However, such deterministic robust stability cannot be ensured only with the
 $Z$ inequalities in (\ref{eq:poly-martingale-expo}), and a more
 severe condition consisting of $Z^2$ inequalities is required in
 association with the parameter-dependent Lyapunov matrix
 (for details, see \cite{Daafouz-SCL01}).
Hence, Theorem~\ref{th:poly-martingale} is not covered by this approach.
\end{remark}

\section{Conclusion}
\label{sc:concl}
In this paper, we studied second-moment stability of discrete-time linear systems with general
stochastic dynamics.
We first showed relations of several notions of second-moment stability,
discussed the time-invariance property of the systems,
and then, derived two types of Lyapunov inequalities characterizing second-moment
exponential stability.
One of them was derived for the systems with the most general
stochastic dynamics (i.e., only under Assumption~\ref{as:bound}, which
is a minimal requirement for defining second-moment stability), and the
other was derived for the systems with essentially bounded random
coefficient matrices (i.e., under Assumption~\ref{as:ess-bound}).
By introducing additional assumptions on the systems, our results can
readily lead us to stability conditions for the corresponding special
cases.
As a demonstration of their usefulness, we provided three applications in
which temporally-independent processes, Markov processes and
polytopic martingales were dealt with for determining system dynamics.
Discrete-time linear systems with any type of stochastic dynamics can be
dealt with in our framework as in those applications.
That is, our framework can unify all the results about second-moment
stability of discrete-time linear systems with stochastic dynamics.
Providing such a framework is expected to facilitate the studies
on analysis and control of stochastic systems drastically.
Although only stability conditions were discussed in this paper, the
techniques used there are considered to be useful also for discussing 
control performance such as $H_2$ and $H_\infty$ norms (provided that
those are appropriately defined in the stochastic sense).


%

%
%
%
%
%

\ifCLASSOPTIONcaptionsoff
  \newpage
\fi



\bibliographystyle{IEEEtran}
\bibliography{ms}

\vfill


\end{document}